\documentclass[prb,twocolumn,showpacs,superscriptaddress]{revtex4-1}
\usepackage[english]{babel}
\usepackage[dvips]{graphicx}
\usepackage{color}
\usepackage{latexsym}
\usepackage{amsmath}
\usepackage{amsthm}
\usepackage{amsfonts}
\usepackage{amssymb}
\usepackage{romannum}

\newcommand{\eps}{\varepsilon}

\begin{document}

\pagenumbering{arabic}

\title{Finite frequency current noise in the Holstein model}
\author{P. Stadler}
\affiliation{Fachbereich Physik, Universit{\"a}t Konstanz, D-78457 Konstanz, Germany}
\author{G. Rastelli}
\affiliation{Zukunftskolleg, Universit{\"a}t Konstanz, D-78457, Konstanz, Germany}
\affiliation{Fachbereich Physik, Universit{\"a}t Konstanz, D-78457 Konstanz, Germany}
\author{W. Belzig}
\affiliation{Fachbereich Physik, Universit{\"a}t Konstanz, D-78457 Konstanz, Germany}
\begin{abstract}
We investigate the effects of local vibrational excitations in the nonsymmetrized current noise $S(\omega)$ of a nanojunction.
For this purpose, we analyze a simple model  -  the Holstein model -
in which the junction is described by a single electronic
level that is coupled to two metallic leads and to a single vibrational mode. 
Using the Keldysh Green's function technique, we calculate the nonsymmetrized current noise  
to the leading order in the charge-vibration interaction. 
For the noise associated to the latter,  we identify distinct terms
corresponding to the mean-field noise and the vertex correction. 
The mean-field result can be further divided into an elastic correction  to the noise  
and in an inelastic correction, the second one being related to energy exchange with the vibration.
To illustrate the general behavior  of the noise induced by the charge-vibration interaction, 
we consider two limit cases.
In the first case, we assume a strong coupling of the dot to the leads with an energy-independent
transmission whereas in the second case we assume a weak  tunneling coupling between the dot and the leads 
such that the transport occurs through a sharp resonant level. 
We find that the noise associated to the vibration-charge interaction shows a complex pattern as a function of the 
frequency $\omega$ and of the transmission function or of the dot's energy level.
Several transitions from enhancement to suppression of the noise occurs in different regions, 
which are determined, in particular, by the vibrational frequency. 
Remarkably, in the regime of an energy-independent transmission,
the zero order elastic noise vanishes at perfect transmission and at positive frequency  
whereas the noise related to the charge-vibration interaction remains finite enabling  
the analysis of the pure vibrational-induced current noise.
\end{abstract}
%
%
%
%
%
%
%
%
%
%
%
\date{\today}
\maketitle
%
%
%

%
%
%
%
\section{Introduction}
\label{sec:intro}

The measurement of fluctuations in macroscopic observables provides information about the microscopic 
dynamics not accessible by the measurement of averaged quantities as for instance the 
charge current [\onlinecite{Blanter2000,Martin:2005,Landauer:1998}].

In quantum devices, different sources contribute to these fluctuations. 
At non-zero temperature, thermal noise causes the fluctuations of the occupation number of energy levels forming the spectrum.
However, the thermal noise is directly related to the conductance via the fluctuation-dissipation theorem 
and hence its measurement  contains equivalent information related the 
conductance [\onlinecite{Callen:1951be}]. 
The situation changes when a voltage is applied to the quantum device.
Then the charge current is in principle time-dependent due to the discreteness of the 
charge [\onlinecite{Blanter2000},\onlinecite{Nazarov:2009wt}] unavoidably appearing in nanoscale conductors. 
For example, an interesting quantity is the zero-frequency noise or shot noise which has been useful for 
a wide range of phenomena [\onlinecite{Blanter2000},\onlinecite{Nazarov:2003}]. 
The shot noise has been employed to reveal the transmission channels in molecular junctions [\onlinecite{Karimi:2016kd,Djukic:2006fv,Tal:2008ct,Kiguchi:2008ec}] or the unconventional quantum of charge in the fractional quantum Hall phase [\onlinecite{Jain:1989,Kane:1994,Saminadayar:1997dw}].

More generally, the quantum nature of the charge current constitute a fundamental source of fluctuations.
which manifests in a finite-frequency current noise $S(\omega)$. 
Radiation (photons) is produced by charge fluctuations and, indeed,
the current noise can be related to the photon exchange.
As a quantum property, $S(\omega)$  can be associated 
to the rate of emission and absorption of photons at the frequency $\omega$ [\onlinecite{Nazarov:2009wt}]. 
The part of the noise at negative frequencies corresponds to the absorption rate of photons by the quantum device  
whereas at positive frequency, the noise  is linked to the rate of emitted photons.
To measure this nonsymmetrized noise, e.g. to distinguish between photon absorption and emission, 
one has to use a quantum detector [\onlinecite{Lesovik:1997iu,Gavish:2000gr,Aguado:2000kv}]. 
Compared to thermal noise, additional information is now contained beyond the one encoded by the conductance.

Molecular-scale devices usually retain their microscopic features, which are then observable in the transport measurements.
Apart from the purely electronic contributions, other degrees of freedom such as vibrational modes can be accessed by spectroscopy. 
In this context, the single-impurity Holstein model [\onlinecite{Glazman:1988,Wingreen:1989,KoenigPRL1996a,Boese:2001,Lundin:2002bt,Kuo:2002gz,McCarthy:2003cz,Braig:2003cb,Flensberg:2003je,Galperin:2004bs,Mitra:2004ema,Wegewijs:2005hi,Kaat:2005ed}] 
has become the paradigmatic model to discuss the effects of charge-vibration interaction in such systems. 
Here, one assumes  a linear coupling between the electron occupation on the dot and the oscillation's amplitude of a harmonic oscillator 
representing the local vibration.
 This model has been theoretically investigated  in different regimes and theoretical methods [\onlinecite{Koch:2005hk,Koch:2005if,Zazunov:2006ex,Koch:2006vq,Koch:2006hu,Hwang:2007ie,Luffe:2008wu,Tahir:2008jj,Schmidt:2009dm,Avriller:2009gn,EntinWohlman:2009iza,Haupt:2009cu,Haupt:2010iza,Schultz:2010eb,Urban:2010eu,Cavaliere:2010ks,Yar:2011wv,Piovano:2011dt,Novotny:2011bsa,Fang:2011ek,Li:2012ik,Skorobagatko:2012hx,Rastelli:2012dh,Santamore:2013hg,Hartle:2013dt,Knap:2013dc,Walter:2013cp,Agarwalla:2015dt,Schinabeck:2016cv,Sowa:2017ez}], in particular using diagrammatic techniques [\onlinecite{Ryndyk:2006ih,Zazunov:2007hva,Egger:2008hha,Dash:2010cz,Rastelli:2010ioa,Dash:2011ez,Souto:2014cm,Chen:2016he}], a resummation approach [\onlinecite{Mera:2015ga},\onlinecite{Pavlyukh:2017hy}], and numerical and non-perturbative methods [\onlinecite{Muhlbacher:2008kz,Hutzen:2012dn,Eidelstein:2013io,Jovchev:2013ki,Souto:2015ho}].

The finite-frequency current noise in quantum dots has been studied in literature with various theoretical approaches and in different kind of contacts, in particular normal contacts [\onlinecite{Engel:2004,Rothstein:2009cl,Ding:2013hi,Moca:2014,Zamoum:2016bw,Crepieux:2017,Orth:2012el,Jin:2015kq}] ferromagnetic contacts [\onlinecite{Braun:2006gj},\onlinecite{Cottet:2008bc}] and hybrid-superconducting contacts [\onlinecite{Droste:2015cz}]. 
Experimentally, both zero-frequency [\onlinecite{Lefloch:2003fp},\onlinecite{Reydellet:2003jj},\onlinecite{Gabelli:2009}] and finite-frequency measurements [\onlinecite{Ubbelohde:2011cy,Basset:2012ff,Basset:2010,Onac:2006bx,Ferrier:2017}] have been reported.
The case of shot noise of a quantum dot
interacting with a local vibration interacting, however, 
has been less investigated, with few exceptions, as for instance in the experiment of Ref.~[\onlinecite{Kumar:2012gm}] 
that reported evidences of shot noise correction associated to the local vibration.

\begin{figure}[t!]			
  	\includegraphics[width=0.8\linewidth,angle=0.]{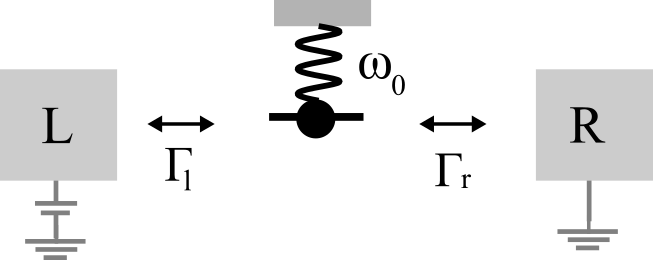} 
  	\caption{Sketch of a quantum dot coupled to a oscillator of frequency $\omega_0$. 
	The dot's energy level is $\varepsilon_0$. The quantum dot is in contacted with two normal 
	contacts with the tunneling rates $ \Gamma_l$ and $\Gamma_r$, respectively. }  
  	\label{fig:NQDS}
\end{figure}

In this work, we discuss the nonsymmetrized finite frequency current noise  $S(\omega)$ 
of a quantum dot with charge-vibration interaction, viz. the Holstein model, 
and with the dot in contact with two normal leads, see Fig.\ref{fig:NQDS}.
$S(\omega)$  encodes the information about the possibility to absorb or emit a photon by the whole 
systems, formed by the mesoscopic conductor (the quantum dot) and the local vibration.
In our approach we consider the weak coupling regime and perform a  perturbative expansion 
in the charge-vibration coupling $\lambda$, viz. $S(\omega) = S_0(\omega) +S_1(\omega)$ with $S_1(\omega)$ 
scaling as $\lambda^2$.
Although we derive a general formula for $S_1(\omega)$,  we focus to two limit cases.
In the first case, we assume  an energy-independent transmission $T$, valid in the limit of strong tunneling coupling or open dot regime. 
In the second case we assume a weak  tunneling coupling between the dot and the leads 
such that the transport occurs through a sharp resonant level with the dot's energy level $\varepsilon_0$.
We analyse $S_1(\omega)$ as a function of $T$ or of $\varepsilon_0$, for the two cases,
and in different regimes of the vibrational frequency, e.i. $\omega_0>eV$ or $\omega_0<eV$, with $eV$ the bias voltage.

%
%
%
%
%
%

For the first limit case, we find several transitions from enhancement to suppression of the noise, occuring
in different frequency ranges.
In particular, at fixed frequency $\omega$, $S_1(\omega)$ shows a non-monotonic behavior as a function of $T$,
with  $S_1(\omega)>0$  at small  or large transmission and  negative values 
$S_1(\omega)<0$ in the intermediate range. 
Remarkably, for $\omega>0$, the zero order elastic noise $S_0(\omega)$ vanishes at $T\simeq1$  
whereas $S_1(\omega)$ is finite enabling the possibility to investigate the pure vibrational-induced 
current noise.
For the second limit case, we find qualitatively similar behaviors although the relevant 
contribution of  $S_1(\omega)$ is strongly localized  in correspondence of characteristic lines 
which are associated to the resonant transport regime. Interestingly, at fixed frequency $\omega$,  
vibrational side bands appear as varying the dot's level $\varepsilon_0$. 
These bands, with spacing equal to the vibrational frequency $\omega_0$,  provide a clear signature of 
the charge-vibration interaction in the noise spectrum.

The paper is organized as follows. 
In Sec.~\ref{sec:model}, we introduce the model and the Keldysh Green's function technique.
Then, in Sec.~\ref{sec:nonsymnoise} we derive the nonsymmetrized current noise $S_1(\omega)$ 
and analyse the different kind of corrections according the technical diagrammatic approach. 
Section~\ref{sec:nonsymnoise} contains the main result of the manuscript, the general formula for $S_1(\omega)$ of the Holstein model.  
In Sec.~\ref{sec:constanttransmission} and \ref{sec:energytransmission} we discuss 
the two limit cases,  the energy-independent transmission regime and the resonant transport, respectively. 
Sec.~\ref{sec:conclusions} contains the final remarks.

%
%
%
%
\section{Model}
\label{sec:model}
In this section, we discuss a model of a quantum dot between conducting contacts (leads) as shown in Fig.~\ref{fig:NQDS}.

We consider the electrons on the quantum dot occupying a spinless state with energy $\varepsilon_0$ and the annihilation and creation operators $\hat{d}^{\phantom{g}}_{}$ and $\hat{d}^\dagger_{}$ on the quantum dot. 
The occupation number reads 
$\hat{n}_d = \hat{d}_{}^{\dagger}\hat{d}_{}^{\phantom{g}}$.
A single harmonic mode of the oscillator with frequency $\omega_0$ has the bosonic annihilation and creation operators $\hat{b}^{\phantom{g}}$  and $\hat{b}^{\dagger}$. 
We assume an interaction between the charge and the amplitude of the oscillator with coupling strength $\lambda$. 
The full Hamiltonian is given by
\begin{equation}
\hat{H} = 
\hat{H}_l + \hat{H}_r + \hat{H}_t+ \varepsilon_0 \hat{n}_d  
+  \lambda \hat{n}_d (  \hat{b}^{\dagger} +\hat{b}^{\phantom{\dagger}}  )
+ \hbar \omega_0  \hat{b}^\dagger \hat{b}^{\phantom{\dagger}}   \, ,
\label{eq:H}
\end{equation}
with the last two terms describing the charge-vibration coupling and the oscillator.
The Hamiltonians of the lateral leads are given by
$\hat{H}_l = \sum_{k} \xi_{l,k}^{\phantom{\dagger}} \hat{c}^\dagger_{k}  \hat{c}^{\phantom{\dagger}}_{k} $
and
$\hat{H}_r = \sum_{k} \xi_{r,k}^{\phantom{\dagger}} \hat{a}^\dagger_{k}  \hat{a}^{\phantom{\dagger}}_{k} $
with the energy $\xi_{\alpha,k}=\varepsilon_k- \mu_{\alpha}$ with $\alpha=(l,r)$ referring to the chemical potential.
The annihilation operators are given 
by  $\hat{c}_{k}$ for the left and $\hat{a}_{k}$ for the right lead. 
The tunneling Hamiltonian is 
$
\hat{H}_t = 
\sum_{k} (
t_{l}^{\phantom{\dagger}} \hat{c}^\dagger_{k} \hat{d}^{\phantom{\dagger}}_{}
+ 
t_{r}^{\phantom{\dagger}} \hat{a}^\dagger_{k} \hat{d}^{\phantom{\dagger}}_{} + \mathrm{H.c.} )
$
with the tunneling amplitudes between the leads and the dot $t_{l}$ and $t_{r}$. 
In the rest of our analysis we assume the wide-band approximation and consider  
the tunnel rates between the quantum dot with the normal contacts 
$\Gamma_l = \pi {\left| t_{l} \right|}^2 \rho_l$ and 
$\Gamma_r = \pi {\left| t_{r} \right|}^2 \rho_r$, respectively, with $\rho_l$ and $\rho_r$ 
the electron density of states at the Fermi level of the leads.

\subsection{Electron Green's functions without charge-vibration interaction $(\lambda=0)$}
\label{subsec:Gdot}
In this subsection we recall the exact results for the electron
Green's functions on the dot without charge-vibration interaction.
These Green's functions denoted by $G(\tau,\tau^\prime)$ constitute the building blocks by which we can express the frequency-dependent current noise in presence of the charge-vibration interaction. 

Since we are interested in the nonequilibrium properties, we defined the contour-ordered Green's functions on the quantum dot as $G(\tau,\tau^\prime)=-i \langle \mathcal{T}_c \hat{d}(\tau) \hat{d}^\dagger(\tau^\prime)\rangle$ with the times $\tau$ and $\tau^\prime$ on the Keldysh contour and the contour-ordering operator $\mathcal{T}_c$ [\onlinecite{Rammer:2007},\onlinecite{Cuevas-Scheer:2010}]. 
We then transform the contour-ordered Green's functions to the real time and define the electron Green's function in Keldysh space as
\begin{equation}
\hat{G}(t,t^\prime) = 
\begin{pmatrix} {G}^{11}(t,t^\prime) & {G}^{12}(t,t^\prime) \\ {G}^{21}(t,t^\prime) & {G}^{22}(t,t^\prime) \end{pmatrix} \, .
\label{eq:G_el}
\end{equation}
with the elements of the matrix defined as
$
{G}^{11}(t,t^{\prime})=-i \langle {\mathcal{T}} \hat{d}(t)\hat{d}^\dagger(t^{\prime}) \rangle 
$,
$
{G}^{22}(t,t^{\prime}) =-i \langle \tilde{\mathcal{T}} \hat{d}(t)\hat{d}^\dagger(t^{\prime}) \rangle   
$,
$
{G}^{12}(t,t^{\prime}) =i \langle \hat{d}^\dagger(t^\prime)  \hat{d}(t^{})  \rangle  
$, and
$
{G}^{21}(t,t^{\prime}) =-i\langle \hat{d}(t)  \hat{d}^\dagger(t^{\prime})  \rangle 
$.
In the above expression, the upper indexes $1$ or $2$ refer to the position of the times $t$ and $t^\prime$ on 
the Keldysh contour  [\onlinecite{Rammer:2007},\onlinecite{Cuevas-Scheer:2010}]. 
The real time-ordering and anti-time ordering operators are denoted by $\mathcal{T}$ and $\tilde{\mathcal{T}}$, respectively. 
In addition to the Green's function in Eq.~\eqref{eq:G_el}, we define the retarded and advanced Green's functions as 
${G}^{R}(t,t^\prime)  =-i\theta(t-t^\prime) \langle \{\hat{d}(t) , \hat{d}(t^\prime) \} \rangle$, 
${G}^{A}(t,t^\prime) =i\theta(t^\prime-t) \langle \{\hat{d}(t) , \hat{d}(t^\prime) \} \rangle$, 
with the anti-commutator $\{\,,\,\}$. These Green's functions are related to the ones in Eq.~\eqref{eq:G_el} by $G^{11}(t,t^\prime)=G^{R,A}(t,t^\prime)+G^{12,21}(t,t^\prime)$ and $G^{22}(t,t^\prime)=G^{21,12}(t,t^\prime)-G^{R,A}(t,t^\prime)$ 
(see appendix \ref{app:GFandSE} for further details). 

From the contour-ordered Green's function, one obtains the Dyson equation 
\begin{equation}
G(\tau,\tau^\prime) \mathord= g(\tau,\tau^\prime) +\!\! \sum_{\alpha=l,r} \! \int \!\! d\tau_1 d\tau_2 g(\tau,\tau_1)\Sigma_\alpha(\tau_1,\tau_2)G(\tau_2,\tau^\prime) \, ,
\label{eq:Dyson_Keldysh}
\end{equation}
with $g(\tau,\tau^\prime)$ the unperturbed dot's Green's functions without tunneling between the dot and 
the leads $\hat{H}_t=0$ and 
the self-energies of the left and right leads $\Sigma_\alpha(\tau,\tau^\prime)$ [$\alpha=(l,r)$] respect 
to the tunneling interaction.
Similarly to the Green's functions, the self-energy appearing in Eq.~\eqref{eq:Dyson_Keldysh} in Keldsyh space are defined as
\begin{equation}
\hat{\Sigma}_{\alpha}(t,t^\prime) = 
\begin{pmatrix} {\Sigma}_{\alpha}^{11}(t,t^\prime) & -{\Sigma}_{\alpha}^{12}(t,t^\prime) \\ -{\Sigma}_{\alpha}^{21}(t,t^\prime) & {\Sigma}_{\alpha}^{22}(t,t^\prime) \end{pmatrix} \, .
\label{eq:Sigma_el}
\end{equation}
with a minus sign on the off-diagonal elements corresponding 
to the different position of $t$ and $t^\prime$ on the Keldsyh contour. 
After a Fourier transformation, we obtain for the Green's functions
\begin{align}
G^R(\varepsilon)&=[\varepsilon-\varepsilon_0+i(\Gamma_l+\Gamma_r)]^{-1} \, \\
G^{12,21}(\varepsilon)&=G^R(\varepsilon)(\Sigma_l^{12,21}(\eps)+\Sigma_r^{12,21}(\eps))G^A(\varepsilon)
\end{align}
with $G^R(\varepsilon)={G^A(\varepsilon)}^*$ and the self-energies of the leads 
\begin{align}
\Sigma_\alpha^{12}(\eps) &= 2i\Gamma_\alpha f_\alpha(\varepsilon) \\
\Sigma_\alpha^{21}(\eps) &= -2i\Gamma_\alpha [1-f_\alpha(\varepsilon)]\, , 
\end{align}
and the Fermi function $f_\alpha(\varepsilon)=\{1-\mathrm{exp}[\beta(\eps-\mu_\alpha)]\}^{-1}$ 
with the chemical potential of the left and right lead $\mu_\alpha$ and the inverse of the temperature $\beta$.

\subsection{Phonon Green's functions without charge-vibration interaction $(\lambda=0)$}

In this subsection we recall the unperturbed phonon Green's function for a single harmonic oscillator which will be useful to 
derive the frequency-dependent current noise in the presence of charge-vibration interaction.
We introduce the symmetrized bosonic operator $\hat{A}(t) = \hat{b}^\dagger(t)+\hat{b}(t)$ and define the phonon Green's function as $D(\tau,\tau^{\prime}) = -i \langle \mathcal{T}_c \hat{A}(\tau) \hat{A}^{\dagger}(\tau^\prime) \rangle$ with the time variable $\tau,\tau'$ 
on the Keldysh contour.  
Then, we change from the contour variable $\tau$ to the real time $t$ 
and write the phonon Green's function in the Keldysh space as a matrix 
\begin{equation}
\hat{D}(t,t^\prime) = 
\begin{pmatrix} D^{11}(t,t^\prime) & D^{12}(t,t^\prime) \\ D^{21}(t,t^\prime) & D^{22}(t,t^\prime) \end{pmatrix} \, ,
\end{equation}
with the phonon Green's functions 
$
{D}^{11}(t,t^{\prime})=-i \langle {\mathcal{T}} \hat{A}(t)\hat{A}^\dagger(t^{\prime}) \rangle 
$,
$
{D}^{22}(t,t^{\prime}) =-i \langle \tilde{\mathcal{T}} \hat{A}(t)\hat{A}^\dagger(t^{\prime}) \rangle   
$,
$
 {D}^{12}(t,t^{\prime}) =-i \langle \hat{A}^\dagger(t^\prime)  \hat{A}(t^{})  \rangle  
 $, and
$
{D}^{21}(t,t^{\prime}) =-i\langle \hat{A}(t)  \hat{A}^\dagger(t^{\prime})  \rangle 
$.
The bare phonon Green's functions read
\begin{align}
D^{\mathop{}_{22}^{11}}(\varepsilon) &= \sum_{s=\pm}\left[-i\pi(2 n_B+1)\delta(\varepsilon+s\omega_0)\pm \mathcal{P} \frac{s}{\varepsilon+s\omega_0}\right] \nonumber
\\
D^{\mathop{}_{21}^{12}}(\varepsilon)&=-2\pi i \left[ (n_B+1)\delta(\varepsilon\pm\omega_0) + n_B \delta(\varepsilon\mp \omega_0) \right]
\end{align}
with the Bose distribution function $n_B=n_B(\omega_0)=[\mathrm{exp}(\beta\omega_0)-1]^{-1}$ 
and the Cauchy principal value of the integral denoted by $\mathcal{P}$. 
For simplicity, we concentrate hereafter, in the main text, 
to the case of zero temperature of the vibration $n_B=0$.

\section{Nonsymmetrized current noise}
\label{sec:nonsymnoise}

In this section we summarize the perturbative expansion of the nonsymmetrized current noise $S(\omega)$ respect to the 
charge-vibration interaction. 

We start with the definition of $S(\omega)$ and then perform a perturbative expansion
in the charge-vibration coupling up to the leading order $S_1(\omega) \propto \lambda^2$.
All terms of this expansion are shown in a diagrammatic representation. 
We obtain that the $S_1(\omega)$ correction can be separated into a mean-field noise and a vertex correction.

\subsection{Diagrammatic approach for the noise}
By using the Hamiltonian Eq.~\eqref{eq:H} we derive the current operator which is given, in the Heisenberg picture, 
by the expression
\begin{equation}
\hat{I}_l(t)=\frac{4ei}{\hbar}\sum_k  t_{l} 
\left[   \hat{c}_k^\dagger(t)\hat{d}(t) - \hat{d}^\dagger(t)  \hat{c}_k(t)
\right] \, .
\end{equation}
We define the nonsymmetrized - frequency-dependent - current noise  on the left lead  as 
\begin{equation}
S(\omega) = \int_{-\infty}^{\infty} d(t-t^\prime) e^{-i \omega(t-t^\prime)}
\langle \hat{I}_l((t)\hat{I}_l((t^\prime)\rangle  \, . 
\label{eq:snons}
\end{equation}
The nonsymmetrized current correlator  on the left lead  can be defined in terms of the Keldysh Green's functions 
as 
\begin{align}
\langle \hat{I}_l(t)\hat{I}_l(t^\prime)\rangle &= S^{21}(t,t^\prime)  \hspace{1cm} t>t^\prime  \label{eq:noiselesgtr_1} 
%
%
%
\end{align}
with the real times $t$ and $t^\prime$.
In Eq.~\eqref{eq:noiselesgtr_1}, 
we have introduced the Green's functions defined on the Keldysh contour and 
which is expressed, generally, in terms of the  times  $\tau$ and $\tau^\prime$ on the Keldysh coutour
\begin{equation}
S(\tau,\tau^\prime) = \, \langle \mathcal{T}_c \hat{I}_l(\tau)\hat{I}_l(\tau^\prime)\rangle \, ,
\label{eq:SKeldysh}
\end{equation}
Then, we calculate the current-current correlation function in Eq.~\eqref{eq:SKeldysh}.

Without the charge-vibration coupling, the correlation function $S(\tau,\tau^\prime)$ can be calculated exactly. 
However, to include the interaction to the oscillator, we use a perturbation expansion  in the charge-vibration coupling 
$\lambda$ which allows to use Wick's theorem  and to decompose the final expression in terms of single-particle 
Green's functions. 
As a result we obtain,  
that the correlator can be written in terms of the zero-order term    
$S_0(\tau,\tau^\prime) \propto \lambda^0$ and the leading order correction  $S_1(\tau,\tau^\prime) \propto \lambda^2$, namely
\begin{equation}
S(\tau,\tau^\prime) =S_0(\tau,\tau^\prime)+S_1(\tau,\tau^\prime) \, .
\label{eq:Sfull}
\end{equation}
Using the diagrammatic representation, the leading order correction can be additionally decomposed into a mean-field contribution $S_{\mathrm{mf}}(\tau,\tau^\prime)$ and a vertex correction, $S_{\mathrm{vc}}(\tau,\tau^\prime)$, 
\begin{equation}
S_1(\tau,\tau^\prime)=S_{\mathrm{mf}}(\tau,\tau^\prime) + S_{\mathrm{vc}}(\tau,\tau^\prime) \, .
\end{equation}
Furthermore, the mean-field correction can be divided into an elastic and an inelastic contribution with the latter 
associated to the absorption of a quantum energy $\omega_0$ by the oscillator.  
\begin{equation}
S_{\mathrm{mf}}(\tau,\tau^\prime) 
 = S_{\mathrm{ec}}(\tau,\tau^\prime)  +  S_{\mathrm{in}}(\tau,\tau^\prime) \, .
\end{equation}
In the following, we discuss in detail the different contributions: the zero-order term $S_0(\tau,\tau^\prime)$, 
the two components of the mean-field correction $S_{\mathrm{ec}}(\tau,\tau^\prime)$ and $S_{\mathrm{in}}(\tau,\tau^\prime)$, 
and the vertex correction $S_{\mathrm{vc}}(\tau,\tau^\prime)$.  

\begin{figure}[t!]
	\begin{center}
		\includegraphics[width=0.5\columnwidth,angle=0.]{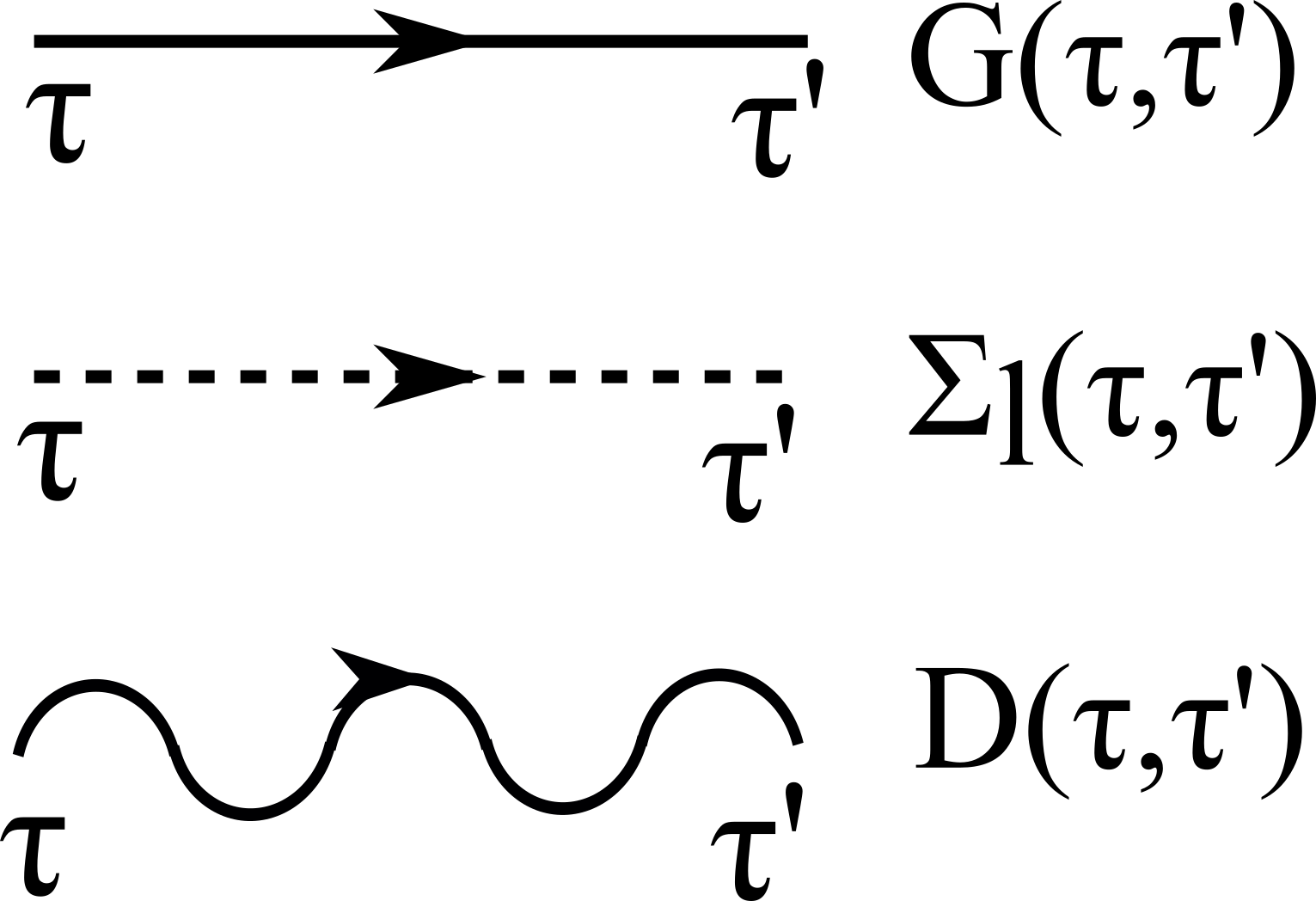}
	\end{center}
	\caption{Elements of Feynman diagrams depending on the Keldysh contour times $\tau$ and $\tau^\prime$.  
	The solid and dashed line represent the Green's function of the dot $G(\tau,\tau^\prime)$ and 
	the self-energy of the left lead $\Sigma_l(\tau,\tau^\prime)$. 
	The phonon propagator $D(\tau,\tau^\prime)$ is denoted by the wiggled line.}    
	\label{fig:NQQD_vib_diagrams}
\end{figure}

Before to proceed, we introduce the notation and the symbol for the elemental block functions, defined 
in the previous section, and appearing in the diagrammatic representation of the noise, see  Fig.~\ref{fig:NQQD_vib_diagrams}. 
The contour-ordered Green's function  of the dot  $G(\tau,\tau^\prime)$ is denoted by a solid line,
the self-energy on the left lead $\Sigma_l(\tau,\tau^\prime)$ is denoted by a  dashed line whereas 
the contour-ordered phonon propagator $D(\tau,\tau^\prime)$ is depicted as a wiggled line. 

\subsection{Zero-order noise  $\lambda=0$}
The zero-order noise for a quantum dot has been studied and well known in the literature.
Here, we simply recall the results using the diagrammatic approach.

\begin{figure}[b!]
	\centering
	\includegraphics[width=0.9\columnwidth,angle=0.]{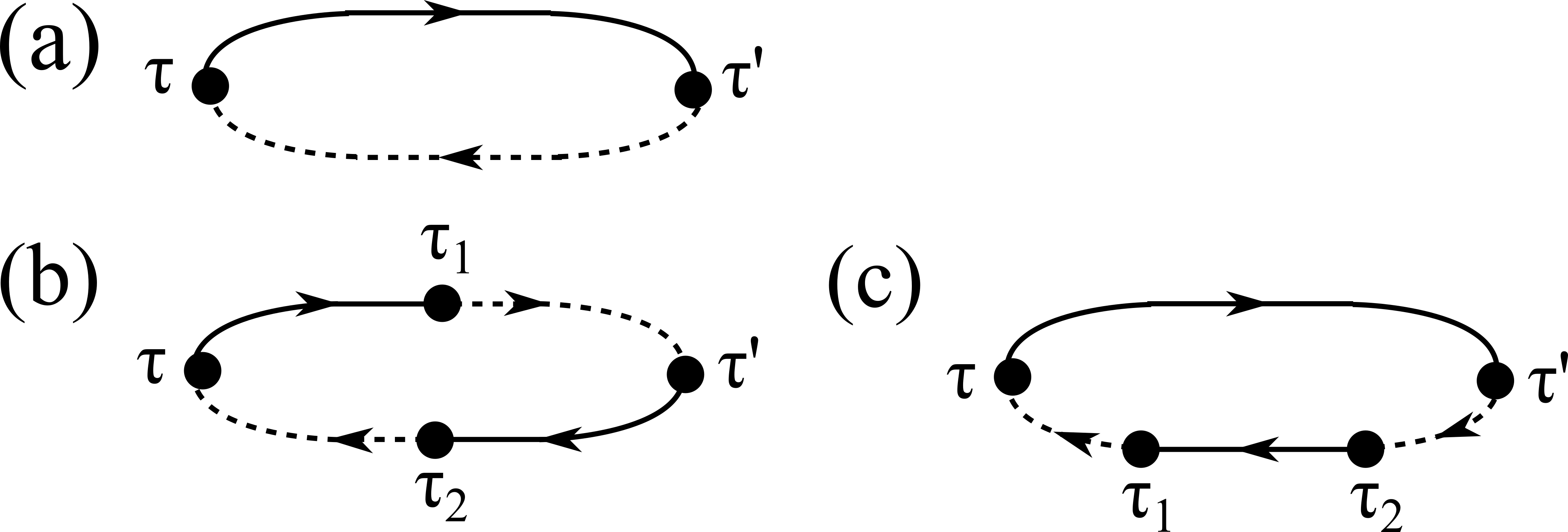}
	\caption{Diagrammatic representation of  $S_0(\tau,\tau^\prime)$ in Eq.~\eqref{eq:S0}. 
	 The solid dot is for the tunneling between the left lead and the dot. 
	 The other symbols are defined in Fig.~\ref{fig:NQQD_vib_diagrams}.}    
	\label{fig:NQQD_S0}
\end{figure}

The diagrammatric representation of the zero-order term $S_0(\tau,\tau^\prime)$  is shown in Fig.~\ref{fig:NQQD_S0} 
and corresponds to 
\begin{align}
S_0(\tau,\tau^\prime) &= \frac{4e^2}{h} \int d\tau_1d\tau_2 \mathrm{Re}\{G(\tau,\tau^\prime) \Sigma_l(\tau^\prime,\tau)  \nonumber  
\\ \nonumber & + 
G(\tau,\tau^\prime)\Sigma_l(\tau^\prime,\tau_1)G(\tau_1,\tau_2)\Sigma_l(\tau_2,\tau)
\\ &+
G(\tau,\tau_2)\Sigma_l(\tau_1,\tau^\prime)G(\tau^\prime,\tau_2)\Sigma_l(\tau_2,\tau^\prime) \} \, ,
\label{eq:S0}
\end{align}
with $G(\tau,\tau^\prime)$ and $\Sigma_l(\tau,\tau^\prime)$ defined in Sec.~\ref{subsec:Gdot}.
The first line in Eq.~\eqref{eq:S0} corresponds to the diagram in Fig.~\ref{fig:NQQD_S0}(a), 
whereas the second and third line correspond to the diagrams in Fig.~\ref{fig:NQQD_S0}(b) and (c), respectively. 
From the expression in Eq.~\eqref{eq:S0} we transform from the contour time to the real time with the definition of the 
matrices $\hat{G}(t,t^\prime)$ and $\hat{\Sigma}(t,t^\prime)$ in Keldysh space given by Eq.~\eqref{eq:G_el} and \eqref{eq:Sigma_el}. 
After the calculations, we obtain the frequency dependent current noise
\begin{align}
S_0&(\omega) \!=\! \frac{e^2}{h} \! \int \! d\eps \{ 
f_r(\eps)(1 - f_l(\eps\!-\!\omega)) T_{lr}(\eps) [1-T_{lr}(\eps\!-\!\omega)]
\nonumber \\  
+&f_l(\eps)(1 - f_r(\eps\!-\!\omega))T_{lr}(\eps\!-\!\omega) [1-T_{lr}(\eps)]
\nonumber \\  
+& f_r(\eps)(1 - f_r(\eps\!-\!\omega))T_{lr}(\eps)T_{lr}(\eps\!-\!\omega)
+[T_{lr}(\eps)T_{lr}(\eps\!-\!\omega)
\nonumber \\ 
+& 4 \Gamma_l^2 \vert G^R(\eps\!-\!\omega)\!-\!G^R(\eps)\vert^2] [f_l(\eps)(1 -f_l(\eps\!-\!\omega))] \}
\label{eq:sf02}
\end{align}
with the transmission function defined as
$
T_{\alpha\beta}(\eps)=4\Gamma_\alpha \Gamma_\beta \vert G^R(\eps) \vert^2 
$ and $(\alpha,\beta)=(l,r)$.

As discussed in Ref.~[\onlinecite{Zamoum:2016bw}], one can identify to each term of Eq.~\eqref{eq:sf02} single 
processes with an absorption or emission of an energy quantum $\omega$, although such processes  
occur on the left lead since we calculated the current-current correlator for $\hat{I}_l(t)$.
Indeed, in Eq.~\eqref{eq:sf02}, all terms are proportional to products of Fermi functions such as $f_\alpha(\eps)(1-f_\beta(\eps-\omega))$ with $(\alpha,\beta)=(l,r)$. 
As an example, the first line in Eq.~\eqref{eq:sf02} describes a process in which an electron from the right lead is transmitted to the left lead 
with the emission of a photon. 
Similar processes can be attributed to the second term and to the third term of the sum inside the integral in 
Eq.~\eqref{eq:sf02}. 
Whereas the first three terms are proportional to the transmission amplitude $\sim T_{lr}$, the last term also contains an additional function $\sim \vert G^R(\eps-\omega)-G^R(\omega)\vert^2$ that represents an interference effect (see Ref.~[\onlinecite{Zamoum:2016bw}]).

\subsection{Mean-field correction}
The corrections proportional to $\lambda^2$ in the perturbation expansion are more involved.
To gain some insight into the finite frequency current noise, we decompose it in several terms 
according the diagrammatic language.

\begin{figure}[h!]
	\centering
	\includegraphics[width=0.85\columnwidth,angle=0.]{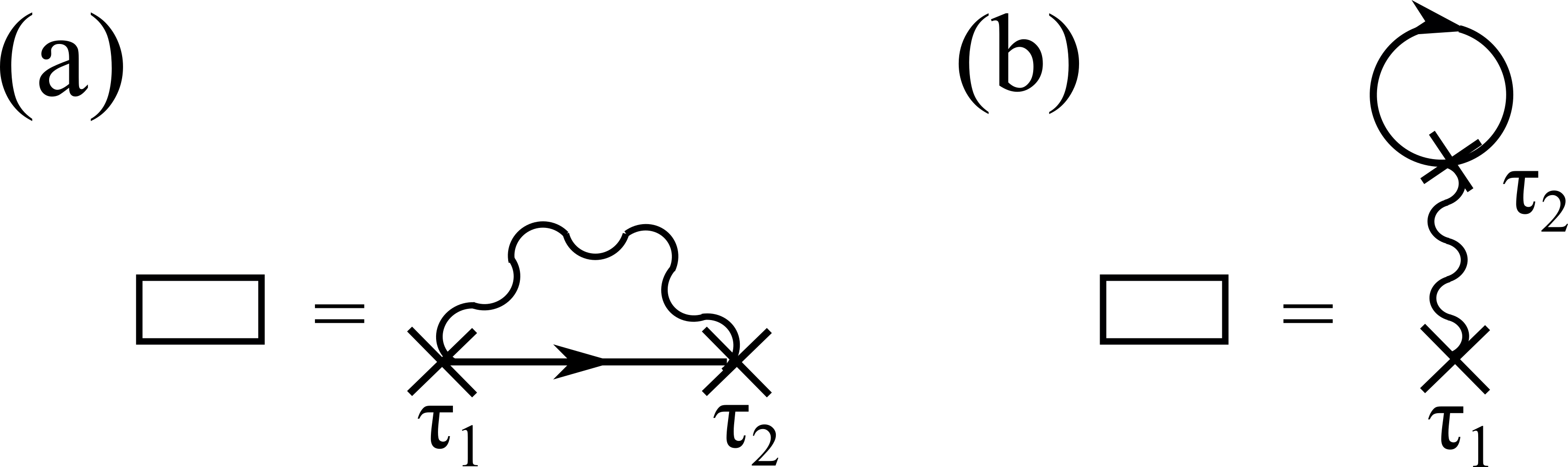}
	\caption{Diagrammatic representation of Eq.~\eqref{eq:rb_tp1} and \eqref{eq:rb_tp2} for the rainbow (a) and tadpole (b) diagrams. 
		The cross represents the vertex due to the charge-vibration interaction.
		The other symbols are defined in Fig.~\ref{fig:NQQD_vib_diagrams}. 
	}    
	\label{fig:NQQD_vib_tprb}
\end{figure}

To discuss the mean-field correction $S_{\mathrm{mf}}(\tau,\tau^\prime)$, 
it is useful to introduce two building block diagrams corresponding 
to the rainbow (tb) and tapole (tp) diagram of the self-energy defined
with respect  the charge-vibration interaction.
In time-domain, these objects can be written as
\begin{align}
\Sigma_{\mathrm{rb}}^{\mathrm{mf}}(\tau_1,\tau_2)  =& i \lambda^2 D(\tau_1,\tau_2) G(\tau_1,\tau_2) \, ,
\label{eq:rb_tp1}
\\
\Sigma_{\mathrm{tp}}^{\mathrm{mf}}(\tau_1,\tau_2) =& - i \lambda^2 D(\tau_1,\tau_2) G(\tau_2,\tau_2)  \, , 
\label{eq:rb_tp2}
\end{align}
and their diagrammatic representation is shown in Fig.~\ref{fig:NQQD_vib_tprb}(a) and (b), respectively. 

\begin{figure}[b!]
	\centering
	\includegraphics[width=0.9\columnwidth,angle=0.]{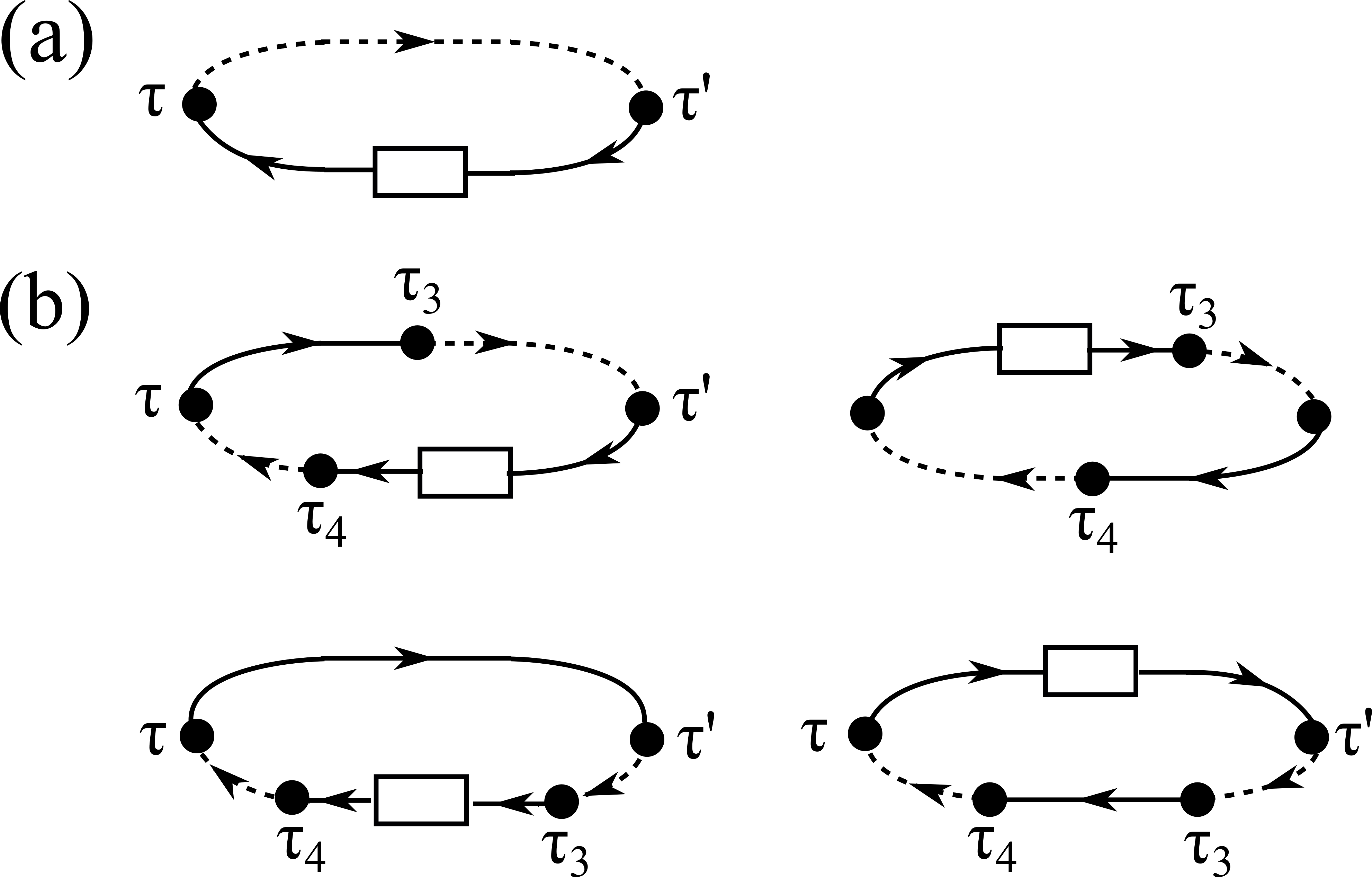}
	\caption{Diagrammatic representation of  $S_{\mathrm{mf}}(\tau,\tau^\prime)$ Eq.~\eqref{eq:meanfieldnoise}. 
	The rectangular box represents either the diagram of Fig.~\ref{fig:NQQD_vib_tprb}(a) or Fig.~\ref{fig:NQQD_vib_tprb}(b). 
	The time arguments $\tau_3$ and $\tau_4$ are internal indexes and the diagrams can be separated into contribution 
	without (a) and with (b) internal indexes. 
	The solid dot is for the tunneling between the left lead and the dot. 
	The other symbols are defined in Fig.~\ref{fig:NQQD_vib_diagrams}.
	 }    
	\label{fig:NQQD_vib_noisetprb}
\end{figure}

With the use of the diagrams in Fig.~\ref{fig:NQQD_vib_tprb}, we can  easily write down 
the contour-time Feynman diagrams contributing to the mean-field noise depicted in Fig.~\ref{fig:NQQD_vib_noisetprb}.
The rectangular box in Fig.~\ref{fig:NQQD_vib_noisetprb} represents either the rainbow or the tadpole diagram of Fig.~\ref{fig:NQQD_vib_tprb}.
The diagrams can be divided into two kinds. 
The first kind of diagrams, shown in Fig.~\ref{fig:NQQD_vib_noisetprb}(a), contain a single self-energy of the left lead, 
two electron Green's functions and the rectangular box. 
The second kind of diagrams [Fig.~\ref{fig:NQQD_vib_noisetprb}(b)] contain two self-energies of the left lead, three electron Green's functions 
and the self-energies due to the electron-vibration interaction. 
There are four different diagrams of the second kind differing by the time labels. 

We can write the mean-field current noise as
\begin{align}
&S_\mathrm{mf}(\tau,\tau^\prime) =  \frac{4e^2}{h} \sum_{\alpha=\mathrm{rb,tp}}  \lbrace
\mathrm{Re}\{{\Sigma}_l(\tau,\tau^\prime)  A_{\mathrm{mf}}^{\alpha}(\tau^\prime,\tau) \} \nonumber
\\ \nonumber &+
   \int d\tau_3 d\tau_4 \mathrm{Re}\lbrace \nonumber
%
%
%
{G}(\tau,\tau_3) {\Sigma}_l(\tau_3,\tau^\prime) 
A_{\mathrm{mf}}^{\alpha}(\tau^\prime,\tau_4){\Sigma}_l(\tau_4,\tau)
\\ \nonumber &+
A_{\mathrm{mf}}^{\alpha}(\tau,\tau_3)  {\Sigma}_l(\tau_3,\tau^\prime) 
{G}(\tau^\prime,\tau_4){\Sigma}_{l}(\tau_4,\tau)
\\ \nonumber &-
{G}(\tau,\tau^\prime)  
{\Sigma}_l(\tau^\prime,\tau_3)A_{\mathrm{mf}}^{\alpha}(\tau_3,\tau_4){\Sigma}_l(\tau_4,\tau)
%
%
\\  &-
A_{\mathrm{mf}}^{\alpha}(\tau,\tau^\prime)
{\Sigma}_l(\tau^\prime,\tau_3){G}(\tau_3,\tau_4){\Sigma}_l(\tau_4,\tau)
\rbrace \rbrace \, .
%
\label{eq:meanfieldnoise}
\end{align}
with the definitions
\begin{align}
A^{\mathrm{rb}}_{\mathrm{mf}}(\tau,\tau^\prime) 
&\mathord= \int d\tau_1 d\tau_2 \, G(\tau,\tau_1) \Sigma^{\mathrm{rb}}_{\mathrm{mf}}(\tau_1,\tau_2) G(\tau_2,\tau^\prime) \, ,
\label{eq:NQDS_vib_ArbtpC} \\ 
A^{\mathrm{tp}}_{\mathrm{mf}}(\tau,\tau^\prime) 
&\mathord= \int d\tau_1 d\tau_2 \, G(\tau,\tau_1) \Sigma^{\mathrm{tp}}_{\mathrm{mf}}(\tau_1,\tau_2) G(\tau_1,\tau^\prime) \, ,
\label{eq:NQDS_vib_Arbtp}
\end{align}
The first term in Eq.~\eqref{eq:meanfieldnoise} corresponds to the diagrams shown in Fig.~\ref{fig:NQQD_vib_noisetprb}(a), 
and the following four terms to the diagrams in Fig.~\ref{fig:NQQD_vib_noisetprb}(b).  

Equation \eqref{eq:meanfieldnoise} is the first main result of the paper 
and corresponds to the mean-field noise in the presence of charge-vibration interaction to the leading order in $\lambda$. 
For this contribution, similar to the current, we can separate  the mean-field correction into an elastic term 
and an inelastic term, i.e. $S_\mathrm{mf}(\tau,\tau^\prime)=S_\mathrm{ec}(\tau,\tau^\prime)+S_\mathrm{in}(\tau,\tau^\prime)$.

\subsubsection{Elastic term of the mean-field correction}
It turns out that the elastic term of the mean-field correction $S_\mathrm{ec}(\tau,\tau^\prime)$ has 
the same structure of the zero-order  current noise in Eq.~\eqref{eq:sf02} 
with a renormalization of the transmission  functions. 
Indeed, it reads 
\begin{widetext}
	\begin{align}
	S_{\mathrm{ec}}(\omega) = -\frac{2e^2}{h} \int d\eps  \,
	&\left\{ f_r(\eps)(1-f_l(\eps-\omega)) [\tilde{T}_{lr}(\eps)-\tilde{T}_{lr}(\eps) T_{lr}(\eps-\omega)-{T}_{lr}(\eps) \tilde{T}_{lr}(\eps-\omega)]
	\right. \nonumber \\ 
	+&f_l(\eps)(1-f_r(\eps-\omega))[\tilde{T}_{lr}(\eps-\omega)-\tilde{T}_{lr}(\eps) T_{lr}(\eps-\omega)-{T}_{lr}(\eps) \tilde{T}_{lr}(\eps-\omega)]
	\nonumber \\
	+& f_r(\eps)(1-f_r(\eps-\omega))
	[\tilde{T}_{lr}(\eps) T_{lr}(\eps-\omega)+{T}_{lr}(\eps) \tilde{T}_{lr}(\eps-\omega)]
	\nonumber \\
	+& f_l(\eps)(1-f_l(\eps-\omega)) [ \tilde{T}_{ll}(\eps)+\tilde{T}_{ll}(\eps-\omega) + T_{lr}(\eps)\tilde{T}_{lr}(\eps-\omega)+\tilde{T}_{lr}(\eps)T_{lr}(\eps-\omega) 
	\nonumber \\
	-&\left. 8\Gamma_l^2 \mathrm{Re}[G^A(\eps)G^R(\eps-\omega)\Sigma^R(\eps-\omega)G^R(\eps-\omega)]
	-8\Gamma_l^2 \mathrm{Re}[G^A(\eps-\omega)G^R(\eps)\Sigma^R(\eps)G^R(\eps)]]  \right\} \, .
	\label{eq:corrsimp}
	\end{align}
\end{widetext}
To formally obtain the result of Eq.~(\ref{eq:corrsimp}), we have to  introduce the {\sl renormalize} transmission 
$\tilde{T}_{\alpha\beta}(\eps) =T_{\alpha\beta}(\eps)\mathrm{Re}[G^R(\eps)\Sigma^R(\eps)]$. 
Then, we can substitute ${T}_{\alpha\beta}(\eps)$ with ${T}_{\alpha\beta}(\eps)+\tilde{T}_{\alpha\beta}(\eps)$  
in Eq.~\eqref{eq:sf02} as well as substitute $G^{R}(\eps)$  with $\rightarrow G^R(\eps)+ G^R(\eps)\Sigma^R(\eps)G^R(\eps)$ 
in the last term of  Eq.~\eqref{eq:sf02}.
Here, $\Sigma^R(\eps)$ is  the retarded self-energy $\Sigma^R(\eps)$
with respect to the charge-vibration interaction
(a summary of the self-energies associated to the charge-vibration interaction is given in Appendix~\ref{app:selfenergies_vib}). 
Then, after these substitutions we keep only the terms proportional to $\lambda^2$. 
In this way, the elastic term to the mean-field correction corresponds to a renormalization of the transmission functions 
and it describes the same processes as the zero-order noise. 
This renormalization can be seen as a virtual exchange of absorption and emission of a phonon from the tunneling charge to the local vibration.

\subsubsection{Inelastic term  of the mean-field correction}
The second term $S_{\mathrm{in}}(\omega) $ appearing in the mean-field corrections 
correspond to the inelastic processes involving the vibration, i.e.  emission or absorption of a 
quantum energy (phonon).

As we focus on the limit of vanishing temperature for the vibration, we have only phonon emission.
In other words, since the frequency-dependent current noise is related to the probability of photon emission or absorption, 
$S_{\mathrm{in}}(\omega)$ is associated to such processes involving not only the tunneling charge 
but even the emission of one vibrational quantum $\omega_0$. 
This term can be written as
\begin{align}
&S_{\mathrm{in}}(\omega) \!=\! \frac{ i e^2}{2h\Gamma_r} \!\!  \int d\eps    T_{lr}(\eps) [f_l(\eps)\Sigma_{21}(\eps)+(1-f_l(\eps))\Sigma_{12}(\eps)]  
\nonumber \\ & \times \sum_s  T_{lr}(\eps+s\omega) [f_l(\eps+s\omega)-f_r(\eps+s\omega)]
\nonumber \\ &+T_{lr}(\eps) \Sigma_{12}(\eps) [1-f_l(\eps-\omega)] 
-T_{lr}(\eps)\Sigma_{21}(\eps)f_l(\eps+\omega) \, .
\label{eq:Sinmf}
\end{align}
with the self-energies $\Sigma_{12}(\eps)$ and $\Sigma_{21}(\eps)$
with respect to the charge-vibration interaction 
defined in the Appendix~\ref{app:selfenergies_vib}. 
Notice that Eq.~\eqref{eq:Sinmf} is real despite the imaginary factor $i$ in front of it.

\subsection{Vertex correction}
\begin{figure}[t!]
 	\begin{center}		
 		\includegraphics[width=0.9\columnwidth,angle=0.]{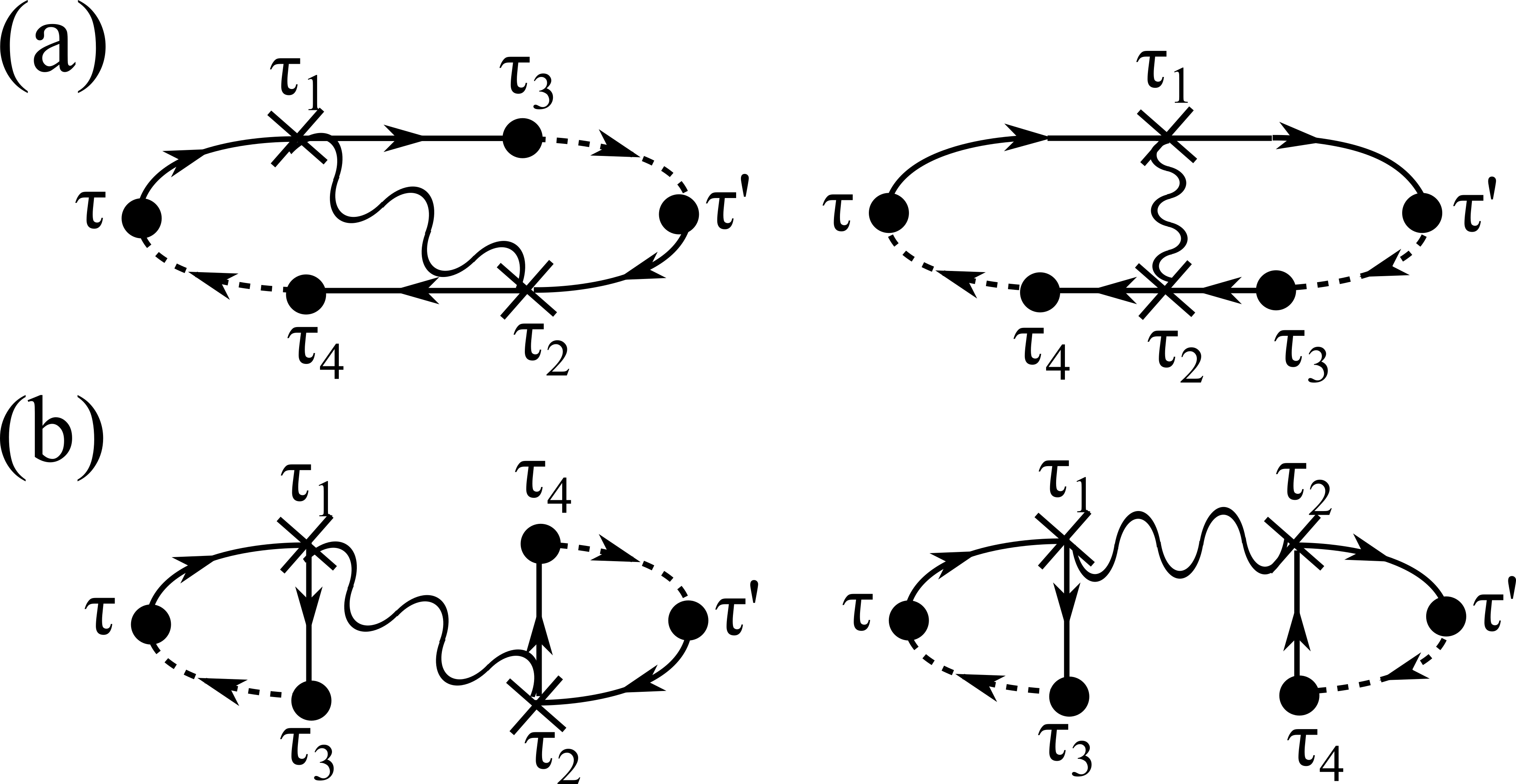}
 	\end{center}
 	\caption{Diagrams corresponding to the vertex correction  $S_{\mathrm{vc}}(\tau,\tau^\prime)$   in Eq.~\eqref{eq:vcrealtime}. 
 		The solid dot is for the tunneling between the left lead and the dot. 
 		The other symbols are defined in Fig.~\ref{fig:NQQD_vib_diagrams}. 
 	}  
 	\label{fig:diagrams1}
 \end{figure}

The vertex correction is more difficult to analyze compared to the mean-field one.
A diagrammatic representation of it is shown in Fig.~\ref{fig:diagrams1} and its 
formula in terms of the times on the Keldysh  contour is given in Eq.~\eqref{eq:vcrealtime}. 
In Appendix~\ref{app:vertex_realtime}, we illustrate the transformation of the Eq.~\eqref{eq:vcrealtime} from the time defined 
on the Keldysh contour $\tau$ to the real time $t$. 

In comparison to the mean-field results in Fig.~\ref{fig:NQQD_vib_noisetprb}, 
the upper and lower branch in Fig.~\ref{fig:diagrams1}(a) are connected due to the interaction with the oscillator. 
The diagrams in Fig.~\ref{fig:diagrams1}(a)  have a similar structure as the rainbow diagrams
whereas  in Fig.~\ref{fig:diagrams1}(b) they a similar structure as the tadpole ones, 
albeit with the external times $\tau$ or $\tau^\prime$. 
\begin{widetext}
	\begin{align}
		S_{\mathrm{vc}}(\tau,\tau^\prime) =i \lambda^2 \int d\tau_1 \, d\tau_2 \, d\tau_3 \, d\tau_4 \, \, \{
		&G(\tau,\tau_1)G(\tau_1,\tau_3)\Sigma_l(\tau_3,\tau^\prime)G(\tau^\prime,\tau_2)G(\tau_2,\tau_4)\Sigma_l(\tau_4,\tau) D(\tau_1,\tau_2)
		\nonumber \\ 
		-&
		G(\tau,\tau_1)G(\tau_1,\tau^\prime)\Sigma_l(\tau^\prime,\tau_3)G(\tau_3,\tau_2)G(\tau_2,\tau_4)\Sigma_l(\tau_4,\tau) D(\tau_1,\tau_2)
		\nonumber \\
		-&
	    G(\tau,\tau_1)G(\tau_1,\tau_3)\Sigma_l(\tau_3,\tau)G(\tau^\prime,\tau_2)G(\tau_2,\tau_4)\Sigma_l(\tau_4,\tau^\prime) D(\tau_1,\tau_2)
		\nonumber \\ 
		+&
		G(\tau,\tau_1)G(\tau_1,\tau_3)\Sigma_l(\tau_3,\tau)\Sigma_l(\tau^\prime,\tau_4)G(\tau_4,\tau_2)G(\tau_2,\tau^\prime) D(\tau_1,\tau_2)\}
		\label{eq:vcrealtime}
	\end{align}
\end{widetext}

\section{Energy independent transmission regime}
\label{sec:constanttransmission}
In principle one can evaluate numerically the different corrections to the noise for an arbitrary range of parameters.
However, in order to gain some insight, we concentrate here on two limit cases that can be worked out analytically.

In the first one, we assume that the energy dependence in the Green's functions can be neglected, 
e.g. we write the retarded Green's functions as $G^R(\eps)=G^R(\eps_F)$, with the energy at the Fermi level $\eps_F$. 
This means in practice that the transmission is energy-independent and we note it as $T=T(\eps_F)$.
This is a good approximation in two cases: (i) when the coupling to the leads is so strong that $\Gamma_l+\Gamma_r \gg \omega_0, eV,\eps_0$ and (ii) when the resonant level is far away from the Fermi energy, i.e. $\eps_0\gg \Gamma_l,\Gamma_r,eV,\omega_0$ 
(we set $\mu_l=\mu_r=0$ in absence of applied voltage).

\subsection{Zero-order current noise $S_0(\omega)$}
It is convenient to start with the discussion of the current noise without the interaction with the oscillator. 
At zero temperature, we can write the elastic current noise as 
\begin{equation}
	S_0^{}(\omega)\!=\!eG
	\begin{cases}
		2 \omega \quad & \omega\!<\!-eV \\
		eV\!-\!\omega\!+\!T (eV\!+\!\omega) & \!-\!eV\!<\!\omega\!<\!0 \\
		(1\!-\!T)(eV\!-\!\omega) & 0\!<\!\omega\!<\!eV \\
		0 & \omega\!>\!eV
	\end{cases}
	\label{eq:S0T}
\end{equation}
with the conductance $G = 2e^2 T/h $. 
$S_0^{}(\omega)$ shows a linear piecewise dependence on the frequency $\omega$ and $eV$.

\begin{figure}[t!]
	\begin{center}
		\includegraphics[width=0.86\columnwidth,angle=0.]{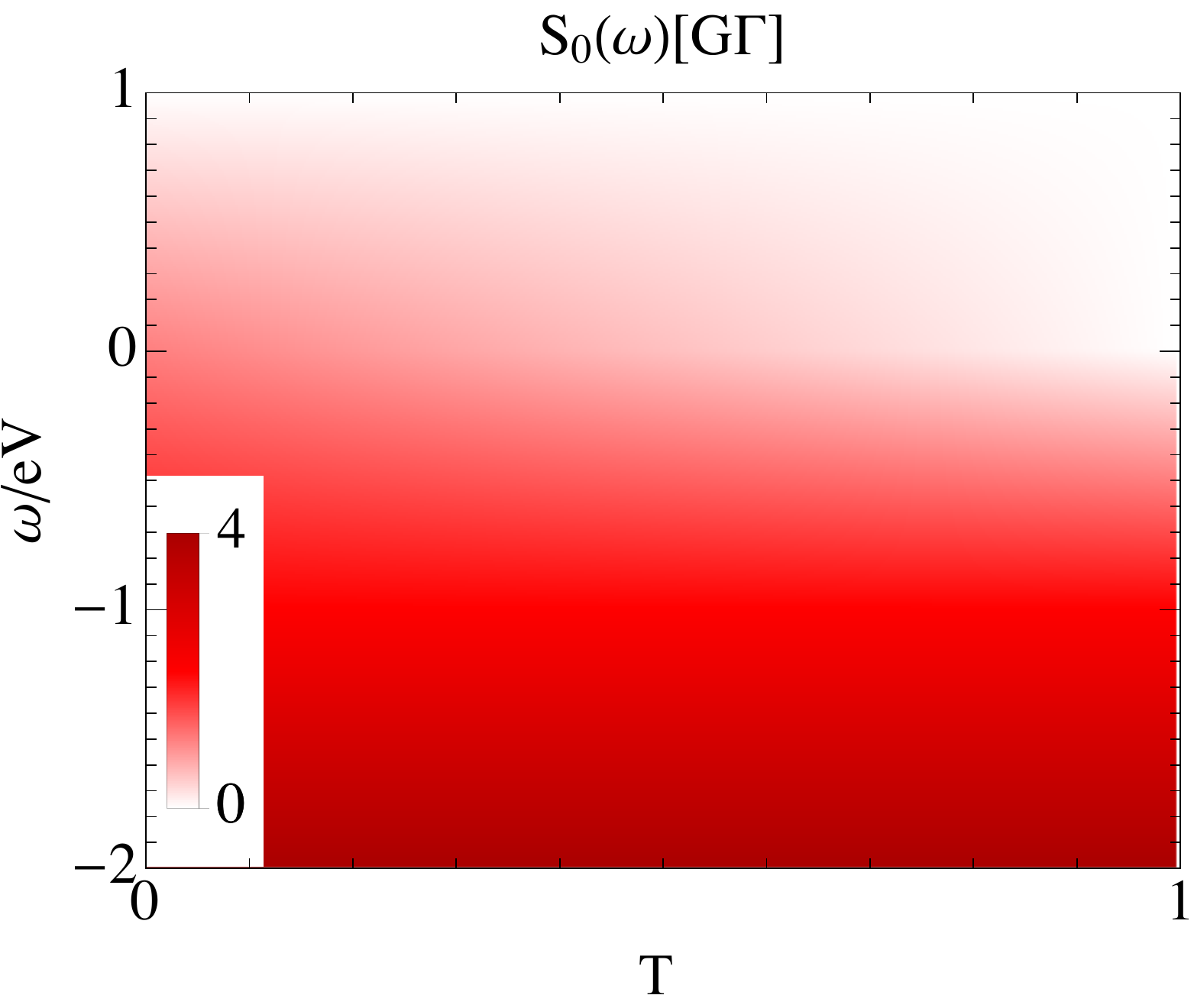}
	\end{center}
	\caption{Zero-order current noise $S_0^{}(\omega)$ as a function of transmission $T$ and noise frequency $\omega$.}  
	\label{fig:S0T}
\end{figure}

Figure \ref{fig:S0T} shows $S_0^{}(\omega)$ as a function of the transmission $T$ 
and the  frequency $\omega$. 
When $\omega>eV$, the noise is zero since a photon due to a single electron tunneling event  here examined - 
can only be emitted with a maximal energy given by the applied bias voltage $\omega=eV$.
When $\omega<-eV$, the current noise scales as $G\omega$ such that in Fig.~\ref{fig:S0T} the current noise seems to be independent of the transmission. 
In the region $0<\omega<eV$, the current noise linearly decreases with increasing transmission and vanishes at $T=1$. 
A similar behavior appears when $-eV<\omega<0$ in which 
the current noise decreases from $eV-\omega$ to $2eV$ by increasing the transmission. 
Finally, for perfect transmission $T=1$, the noise vanished at positive frequency.

\subsection{Correction to the noise $S_1(\omega)$}
In this section we analyze the correction $S_1(\omega)$ to the 
 finite-frequency current noise  due to the  charge-vibration interaction 
in the limit of energy independent transmission of the quantum dot.

\begin{figure}[t!]
	\begin{center}		
		\includegraphics[width=0.86\columnwidth,angle=0.]{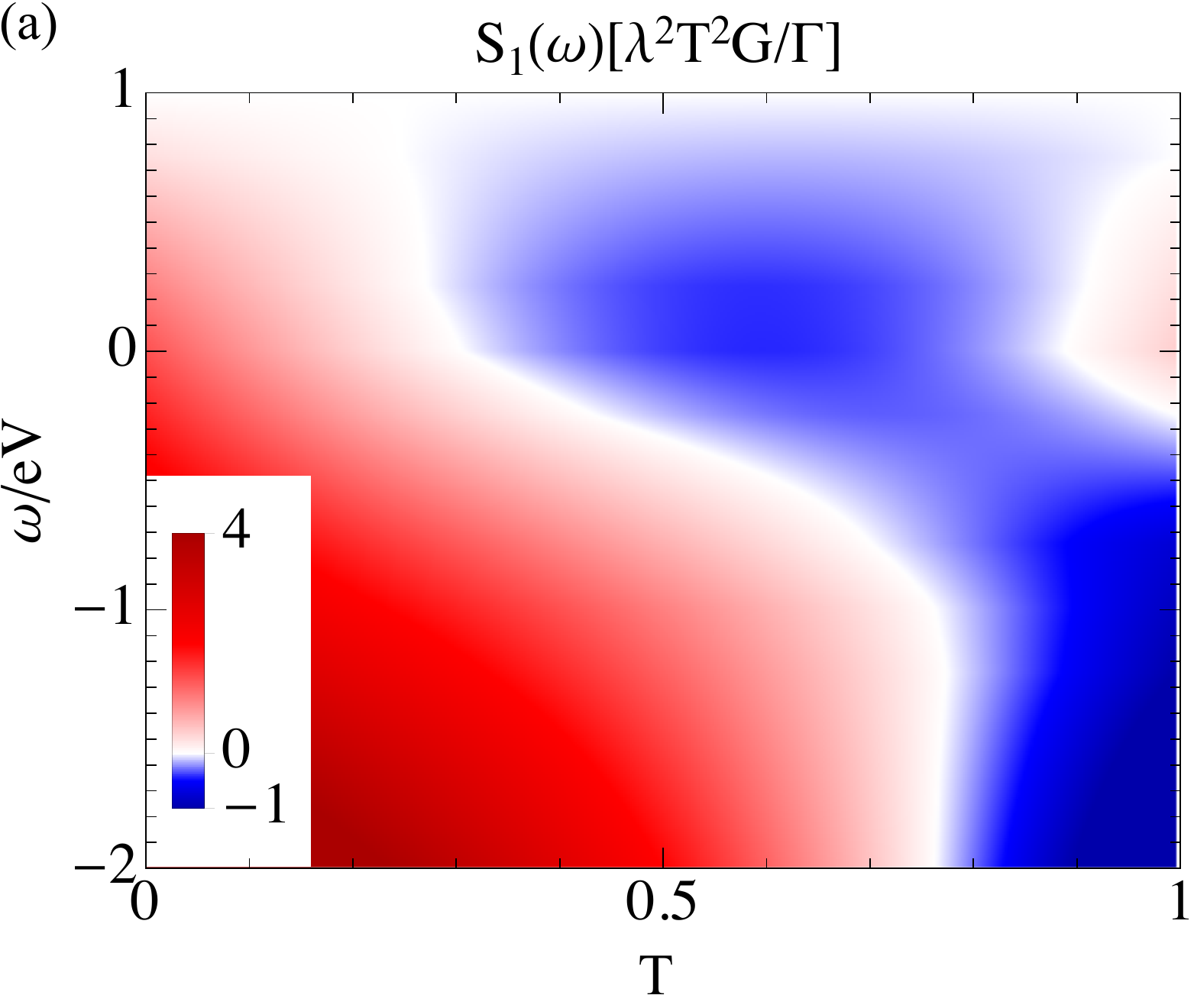}
		\hspace{1cm}
		\includegraphics[width=0.86\columnwidth,angle=0.]{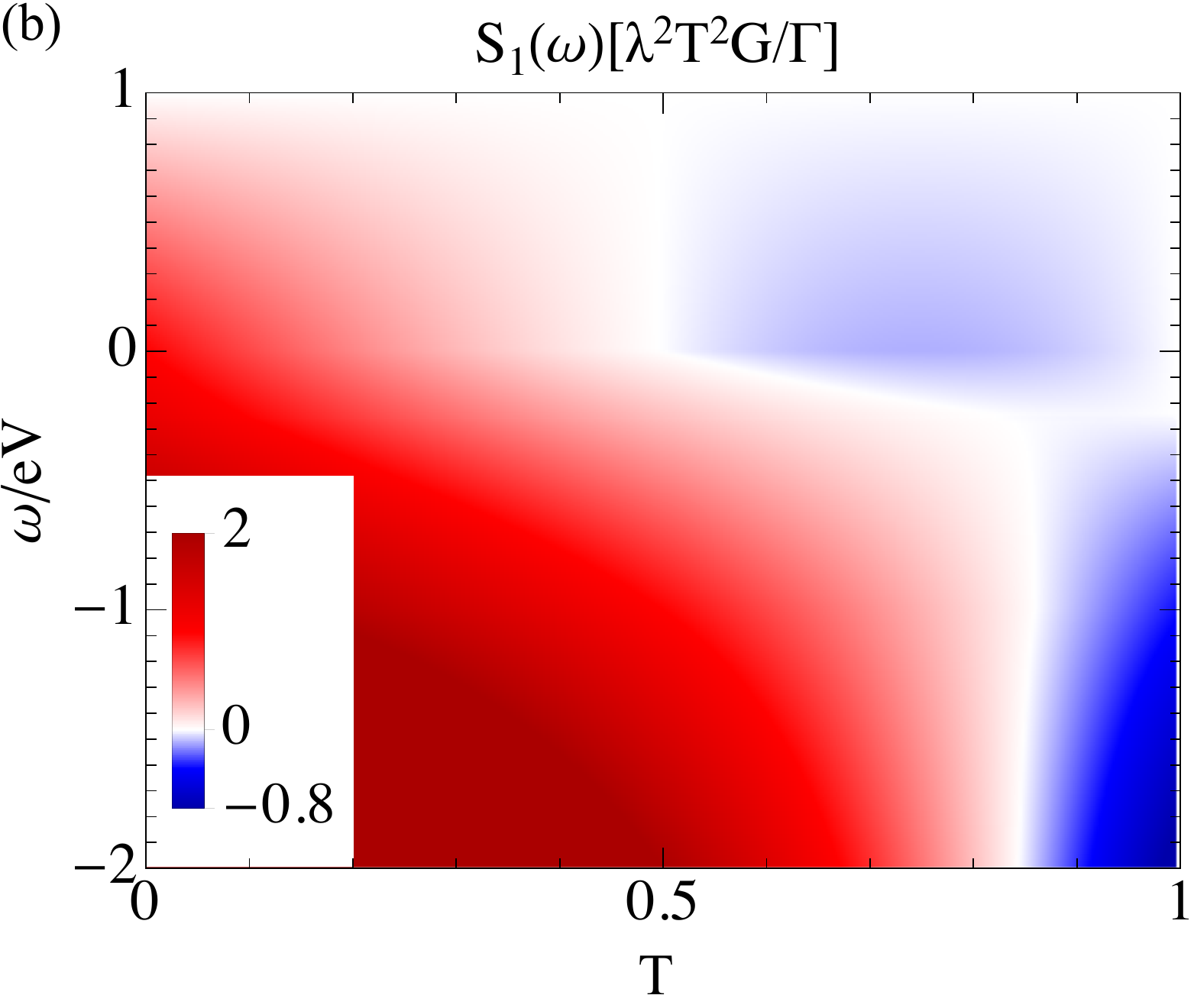}
	\end{center}
	\caption{Correction to the noise $S_1(\omega)$ as a function of transmission $T$ and frequency $\omega$.
	 In (a)  $\omega_0=0.25eV$ and in (b)   $\omega_0=1.25eV$. 
	 In (a) the correction to the noise shows a double transition in the frequency range $-\omega_0<\omega<eV-\omega_0$ 
	  from positive to negative value.  
	   At $T=1$, $S_1(\omega)$ is finite and positive in the frequency range $0<\omega<eV-\omega_0$, in contrast to the zero-order 
	   noise $S_0(\omega)=0$ in Fig.~\ref{fig:S0T}.
	}  
	\label{eq:S_T}
\end{figure}

The charge-vibration interaction leads to a suppression or an enhancement of the current noise: as 
shown in  Fig.~\ref{eq:S_T}: depending on the transmission $T$ and the frequency $\omega$, $S_1(\omega)$ 
can be positive or negative.
Such a complex behavior is caused by the interplay of the mean-field corrections and the vertex correction.
We note that the corrections to the current noise is scaled with $\lambda^2$ and the total noise is positive for all parameter range. 

Although the elastic term $S_{\mathrm{ec}}(\omega)$ and the inelastic term $S_{\mathrm{in}}(\omega)$,
of the mean field correction and the vertex correction $S_{\mathrm{vc}}(\omega)$ 
can be further simplified under the assumption of energy-independent transmissions, 
here we discuss the full expression $S_1(\omega)$ for energy-independent transmission.
The detailed  expressions for the individual terms are given in Appendix \ref{sec:T_independent}. 
Generally, $S_1(\omega)$  has a piecewise linear dependence of the current noise on the frequency $\omega$ 
controlled by the frequency of the oscillator $\omega_0$ and the voltage $eV$. 
When $\omega>eV$, the noise correction $S_1(\omega)$  vanishes in similar way as $S_0(\omega)$, as previously discussed. 
We distinguish two regimes depending on the frequency of the oscillator and the voltage.
The first regime is given by $\omega_0<eV$ in which the voltage is sufficiently large to excite the oscillator. 
In the second one $\omega_0<eV$,  the voltage is smaller than the frequency of the oscillator 
such that an additional absorption of a photon by the electron is needed to excite the oscillator.

For $eV>\omega_0$, the correction to the noise is given by
\begin{widetext}
	\begin{align}
	S^{eV>\omega_0}_{1}(\omega) \!=\! \frac{\lambda^2e^2 T^2}{2\Gamma^2}  \!\! 
	\begin{cases}
	\left(4 T \!-\!3\right)\omega\!+\!(2T\!-\!1)\omega_0 & \omega  \!<\!-eV-\omega_0
	\\
	\left(3 T \!-\! \frac{5}{2}\right)\omega\!+\!(T\!-\!\frac{1}{2})\omega_0 -\left(T-\frac{1}{2}\right)eV& -eV-\omega_0\!<\!\omega \!<\!-eV
	\\
	\left(4 T^2-2T-\frac{3}{2}\right)\omega+(T-\frac{1}{2})\omega_0+\left(4 T^2-6T+\frac{3}{2}\right)eV & -eV\!<\!\omega \!<\!\mathrm{min}(-\omega_0,-eV+\omega_0)
	\\
	\left(2 T^2-\frac{3}{2}\right)\omega-\left(2T^2-3 T+\frac{1}{2}\right)\omega_0+\left(4T^2-6T+
	\frac{3}{2}\right)eV \\ \hspace{1.3cm} - T (eV (1 -2 T) + (3 - 4 T) \omega+ \omega_0) \theta(
	eV - 2 \omega_0)& \mathrm{min}(\!-\!\omega_0,\omega_0\!-\!eV)\!<\!\omega \!<\!\mathrm{max}(\!-\!\omega_0,\omega_0\!-\!eV)
	\\
	\left(4 T^2\!-\!T\!-\!\frac{3}{2}\right)\omega \!-\!(4T^2\!-\!4T\!+\!\frac{1}{2})\omega_0\!+\!\left(6T^2\!-\!7T+
	\frac{3}{2}\right)eV&\mathrm{max}(-\omega_0,\omega_0-eV)\!<\!\omega\!<\!0
	\\
	\left(5T\!-\!4T^2\!-\!\frac{3}{2}\right)\omega \!-\!(4T^2\!-\!4T\!+\!\frac{1}{2})\omega_0\!+\!\left(6T^2\!-\!7T\!+\!
	\frac{3}{2}\right)eV& 0\!<\!\omega\!<\!\mathrm{min}(\omega_0,eV-\omega_0)
	\\
	\left(\!-\!2T^2\!+\!3T\!-\!1\right)\omega\!+\!2 T (1\!-\!T)\omega_0 \!+\!(1\!-\!T)(1\!-\!4 T)eV \\ \hspace{0.8cm} +\!\frac{1}{2} (eV (1 \!-\! 2 T)^2 \!-\! (1\! -\! 8 T (1 \!-\! T))\omega \!-\! \omega_0) \theta(eV \!-\! 2 \omega_0)&\mathrm{min}(\omega_0,eV-\omega_0) \!<\!\omega\!<\!\mathrm{max}(\omega_0,eV-\omega_0) 
	\\
	\left(-4 T^2+5T-1\right)\omega+(1-T)(1-4T)eV & \mathrm{max}(\omega_0,eV-\omega_0) \!<\! \omega \!<\! eV
	\\
	0 & \omega \!>\! eV
	\end{cases}
	\label{eq:SEC_T1}	
	\end{align}
\end{widetext}

For $\omega<eV$, the noise shows a linear dependence as a function of $\omega$ which can be divided into nine frequency intervals. 
The behavior of the noise Eq.~\eqref{eq:SEC_T1} is illustrated  in Fig.~\ref{eq:S_T}(a) 
as a function of the transmission $T$ and  frequency $\omega$ for $\omega_0=0.25eV$. 
The separation of the correction to the noise in intervals appears as a kink in the frequency-dependence, i.~e. 
at $\omega=0,\pm \omega_0,-eV+\omega_0,-eV$ and $\omega_0=-eV-\omega_0$. 
For the other intervals of Eq.~\eqref{eq:SEC_T1}, the change in slope is too small to reveal the kinks in Fig.~\ref{eq:S_T}(a).
The transition from the positive to the negative correction to the emission noise in the interval $eV-\omega_0 < \omega <eV$ appears at $T=0.25$ 
and is independent from the oscillator frequency $\omega_0$ as long as $\omega_0 < eV$, see Eq.~\eqref{eq:SEC_T1}. 
In the interval $-\omega_0 <\omega<eV-\omega_0$, 
the correction to the noise is positive for small transmission, 
becomes negative by increasing the transmission and is again positive when $T\rightarrow 1$. 
Such a double transition as has been obtained for the shot noise ($\omega=0$) in Ref.~[\onlinecite{Haupt:2009cu}] 
and measured in Ref.~[\onlinecite{Kumar:2012gm}].
Here we demonstrate that such a double transition also appears at finite frequencies. 
Finally, when $\omega<-\omega_0$, the correction to the noise has a single transition in which it switches from 
positive to negative values as increasing the transmission $T$.

Interesting features appear in $S_1(\omega)$ for the emission noise ($\omega>0$) 
at perfect transmission $T=1$. 
In this case,  the vertex correction vanishes at positive frequencies.
The noise $S_1(\omega)$ is given only by the mean-field corrections and vanishes when $\omega>eV-\omega_0$ 
due to the cancellation of the elastic correction and the inelastic one,  
as can be seen from Eqs.~\eqref{eq:Sin1T_Appendix} 
and \eqref{eq:Sec1T_Appendix}  in the Appendix.
However, $S_1(\omega)$  is finite and positive when $0<\omega<eV-\omega_0$.
Here, as recalled in the previous section, the zero-order noise  $S_0(\omega)$ vanishes for positive frequencies at $T=1$, e.g. 
it scales linearly to zero as $1-T$.
Hence, a finite emission noise for perfect transmission (or close to it $T\rightarrow 1$)  can be an intrinsic 
signature of the charge-vibration interaction of the quantum transport through the tunnel junction.

We now turn to the discussion of the correction to the noise for $eV<\omega_0$. 
In this case the oscillator cannot be inelastically be excited.
The corrections to the noise can be divided into eight intervals  
 \begin{widetext}
	\begin{align}
	S^{eV<\omega_0}_{1}(\omega) = \frac{\lambda^2e^2}{2h\Gamma^2}  T^2
	\begin{cases}
	(4 T-3) \omega + ( 2 T-1) \omega_0 & \omega  <-eV-\omega_0
	\\
	(3 T -\frac{5}{2}) \omega + (T-\frac{1}{2}) \omega_0 +(\frac{1}{2} - T) eV & -eV-\omega_0<\omega <-\omega_0
	\\
	(3 T-\frac{5}{2}) \omega + ( T-\frac{1}{2}) \omega_0 -eV (T-\frac{1}{2}) & -\omega_0<\omega <-eV
	\\
	(2 T^2-\frac{3}{2} ) \omega + (T-\frac{1}{2}) \omega +(2 T^2 - 4 T +\frac{3}{2}) eV& -eV<\omega <eV-\omega_0
	\\
	-(1 - T) (1 + 2 T) \omega +eV (1 - T) (1 - 2 T) & eV-\omega_0<\omega<0
	\\
	( -2 T^2+3 T-1) \omega + (T-1 ) (2 T-1) eV& 0<\omega<eV
	\\
	0 & \omega > eV \, .
	\end{cases}
	\label{eq:SEC_T2}	
	\end{align}
\end{widetext}

Figure~\ref{fig:S1T}(b) shows the correction to the noise at $\omega_0=1.25eV$. 
In this case the correction to the noise is positive at small transmission and changes sign once time 
by increasing the transmission.  
Interestingly, the sign change in the emission noise $(\omega>0)$ appears at $T=0.5$ independent of $\omega_0$ as long as $\omega_0>eV$. 
%
%
%
%
%
%
%
%
%

\section{Resonant transmission regime}
\label{sec:energytransmission}
In this section, we focus on the regime in which the electrons tunnel through the resonant level at $\varepsilon_0$. 
This regime is characterized by a tunneling coupling much smaller then the voltage and the frequency of the oscillator with tunneling rates $\Gamma_l,\Gamma_r \ll eV,\omega_0$. 
In this case we have to take the energy-dependence of the Green's functions appearing the zero-order 
noise [Eq.~\eqref{eq:sf02}],  the elastic term  [Eq.~\eqref{eq:corrsimp}] and the inelastic term [Eq.~\eqref{eq:Sinmf}] 
of the mean-field correction,  and the vertex correction [Eq.~\eqref{eq:vcrealtime}] into account.
We report here the numerical results of these expressions. 
Similar to the previous section, we first study the zero-order noise $S_0(\omega)$ and then 
we discuss the correction $S_1(\omega)$.

\subsection{Zero-order current noise $S_0(\omega)$}
\label{subsec:S_0}
\begin{figure}[b!]
	\begin{center}
		\includegraphics[width=0.86\columnwidth,angle=0.]{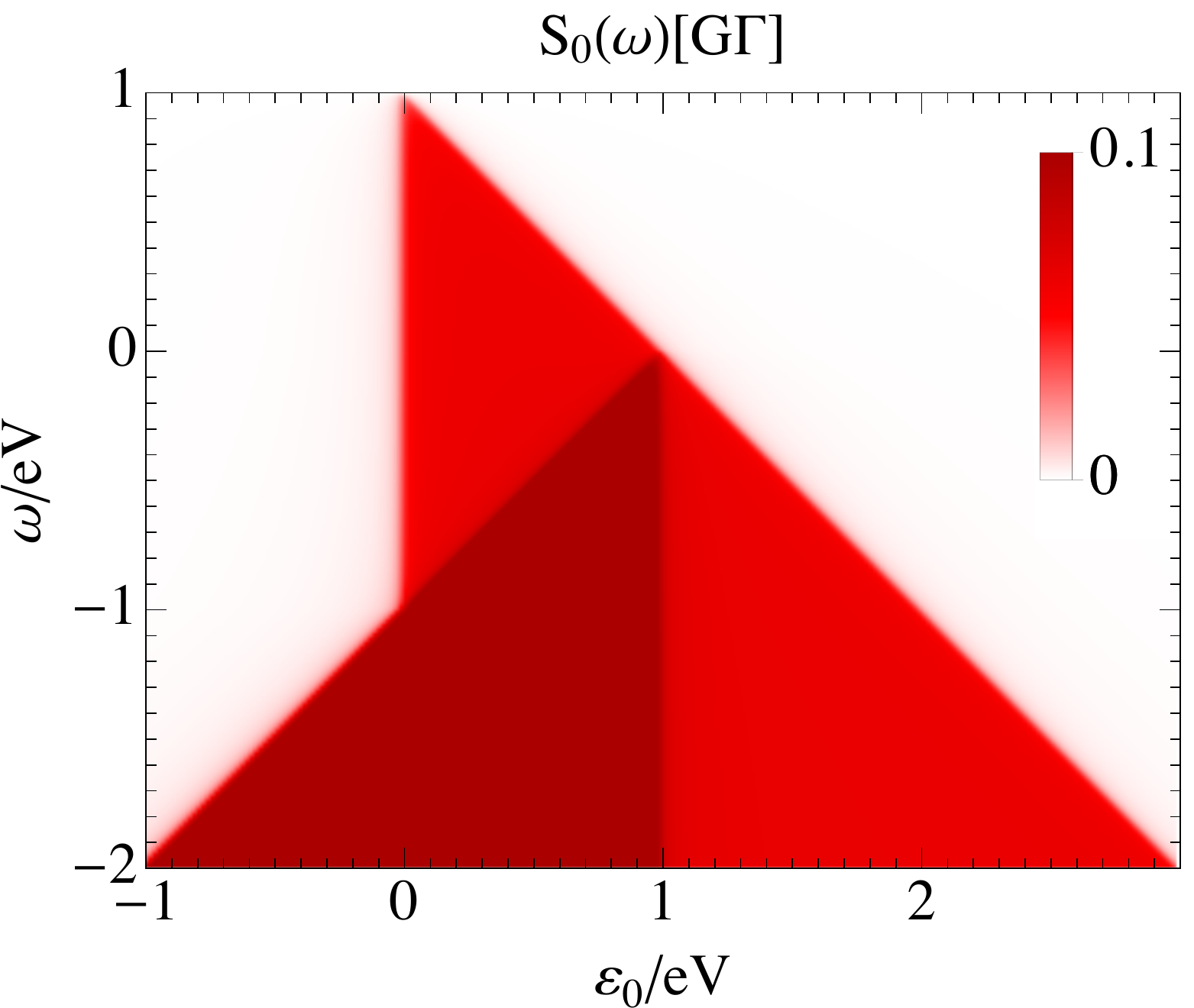}
	\end{center}
	\caption{Current noise $S_{0}(\omega)$ as a function of the dot's energy level $\varepsilon_0$ and noise frequency $\omega$ 
	at zero temperature. The coupling to the leads is symmetric $\Gamma_l=\Gamma_r=\Gamma$ with $\Gamma=0.01eV$.  }  
	\label{fig:S0E}
\end{figure}

Figure \ref{fig:S0E} shows $S_0(\omega)$ as a function of the gate voltage $\varepsilon_0$ and the frequency $\omega$. 
The voltage is applied on the left lead $\mu_l=eV$ and $\mu_r=0$. 
Since we discuss the resonant regime, we fix a small coupling to the leads with $\Gamma_l=\Gamma_r=0.01eV$.

\begin{figure}[t!]
	\begin{center}
		\includegraphics[width=1\columnwidth,angle=0.]{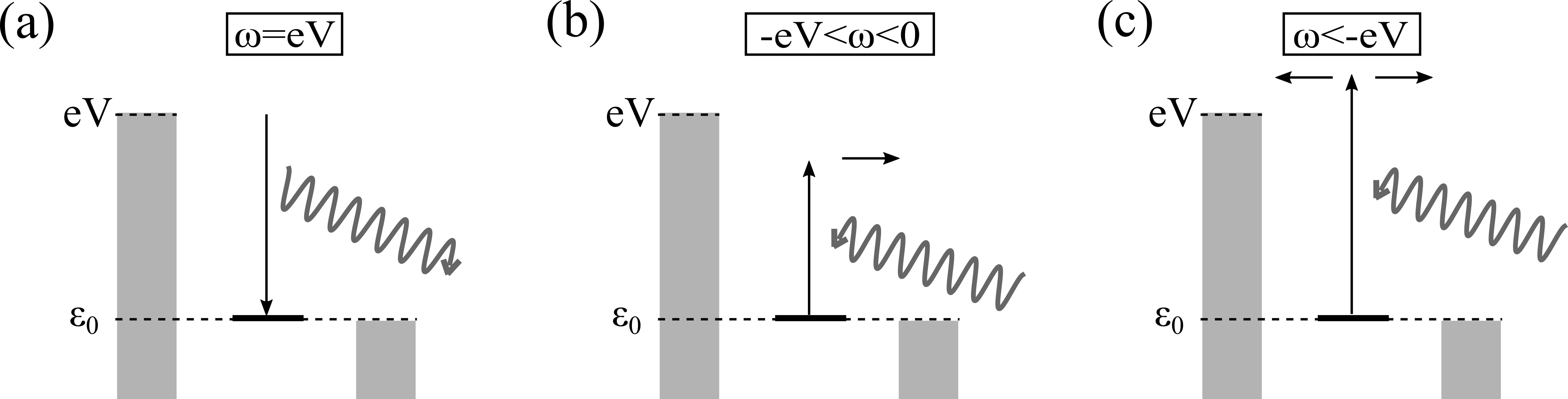}
	\end{center}
	\caption{Schematic of processes corresponding to the emission (a) and absorption (b,c) of a photon for the resonant transport regime 
	at $\varepsilon_0=0$. 
	These processes explain the behavior of the elastic noise in Fig.~\ref{fig:S0E}. 
	In (a), a photon with the maximal frequency $\omega=eV$ is emitted by the contact. 
	In (b) an electron is excited by absorbing a photon with energy smaller than the voltage. 
	After the excitation, the electron can only tunnel to the right lead. 
	In (c), the photon has an energy larger than the voltage such that the excited electron can tunnel to the left and right lead.
	}  
	\label{fig:S0E_sketch}
\end{figure}

First, consider the case of the emission noise $(\omega> 0)$. 
Since the voltage is applied on the left lead, an electron with energy $\varepsilon=\mu_l$ can tunnel from the left chemical potential to the quantum dot at $\varepsilon_0$ and thereby emit a photon with energy $\omega = eV-\varepsilon_0$. 
An example of this process is shown in the inset of Fig.~\ref{fig:S0E_sketch}(a) for $\varepsilon_0=0$.
Then,  when then gate voltage in increased, e.g. $\varepsilon_0=eV/2$,  the maximal energy for the electron to emit a photon reduces to $eV/2$.
In other words, in the resonant transport regime here discussed, the effective, maximal  energy of the photon emitted 
is given by $eV-\varepsilon_0>0$. 
Eventually, when the energy level of the quantum dot is tuned outside the voltage bias region,  $\varepsilon_0>eV$ or $\varepsilon_0<0$,
an electron cannot tunnel through the resonant level and the noise vanishes.

We now consider the absorption noise at negative frequencies ($\omega<0$).
Similar to the emission noise, the effective, minimum amount of absorbed energy by a photon  
is given by $|\varepsilon_0-eV|$ otherwise the electron from the lead lead can not tunnel into the dot's level.
Thus the current (absorption) noise vanishes again for $\varepsilon_0>eV$.

To illustrate the general behavior in this frequency range, we can assume $\varepsilon_0=0$ for simplicity.
When $\vert \omega\vert<eV$, a photon is absorbed by the electrons from the quantum dot's level 
and then tunnel {\sl only} to  the right lead since $\omega<eV$. 
An example of such an absorption process in shown in the right inset of Fig.~\ref{fig:S0E_sketch}(b).
This absorption process with the tunneling to the right lead appears for all frequencies in the range $|\omega|<eV$. 
However, when the frequency is larger than the voltage $|\omega|>eV$, 
an electron from the quantum dot's level at $\varepsilon_0=0$ can, after the absorption of a photon, 
tunnel to the right or the left lead, Fig.~\ref{fig:S0E_sketch}(c) leading to an increase of 
the current noise. 
Similar discussion is valid at finite values of $\varepsilon_0 $, albeit that the electron can also tunnel from the leads to the dot, 
and it explains the step-like increase of the current noise  corresponding to the  dark red region in Fig.~\ref{fig:S0E}.

%
%
%
%

\subsection{Correction to the noise $S_1(\omega)$}
In this section we discuss the correction to the noise $S_1(\omega)$ in the resonant transport regime.

%
%
%
%
\begin{figure}[t!]
	\begin{center}		
		\includegraphics[width=0.86\columnwidth,angle=0.]{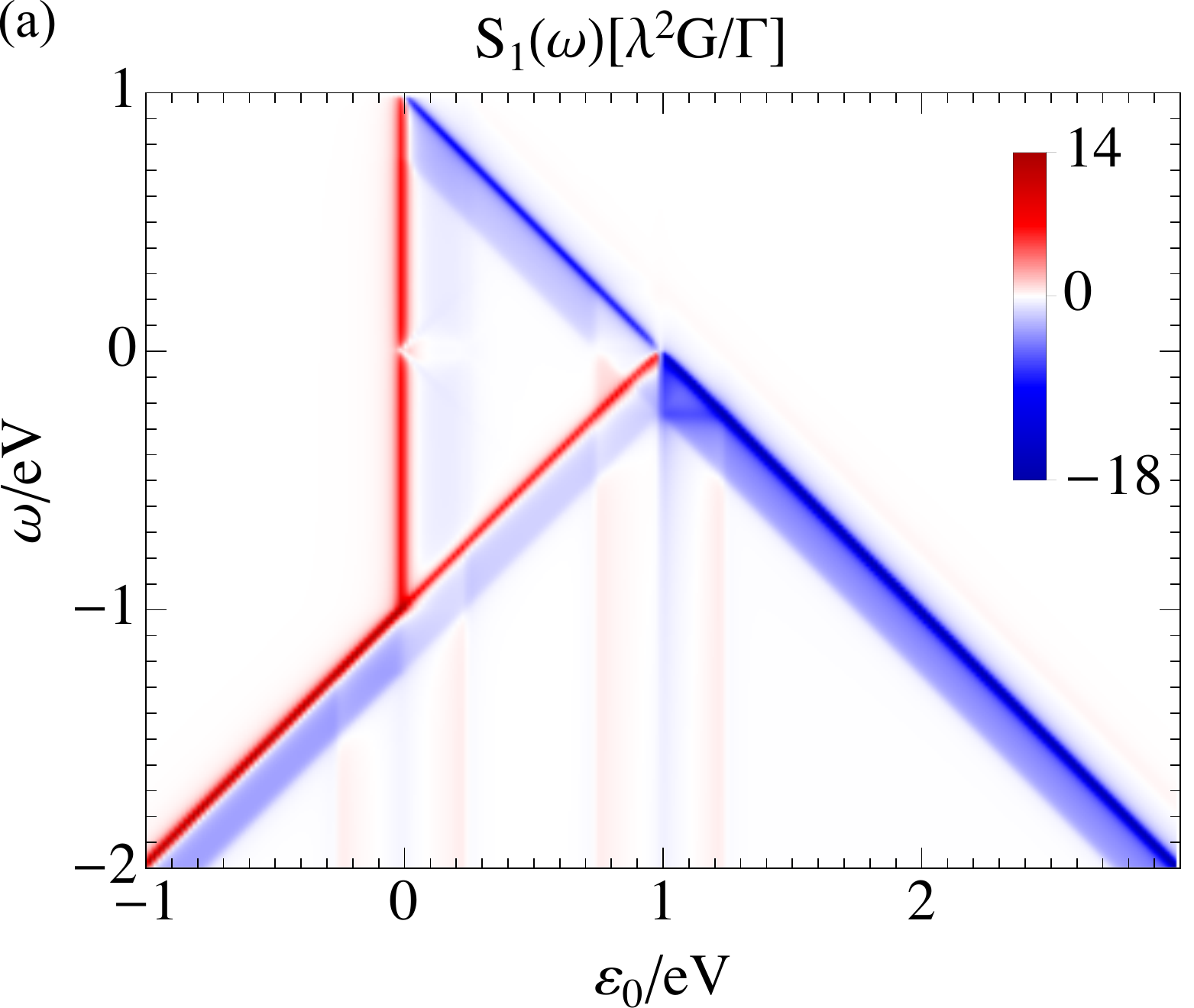}
		\hspace{1cm}
		\includegraphics[width=0.86\columnwidth,angle=0.]{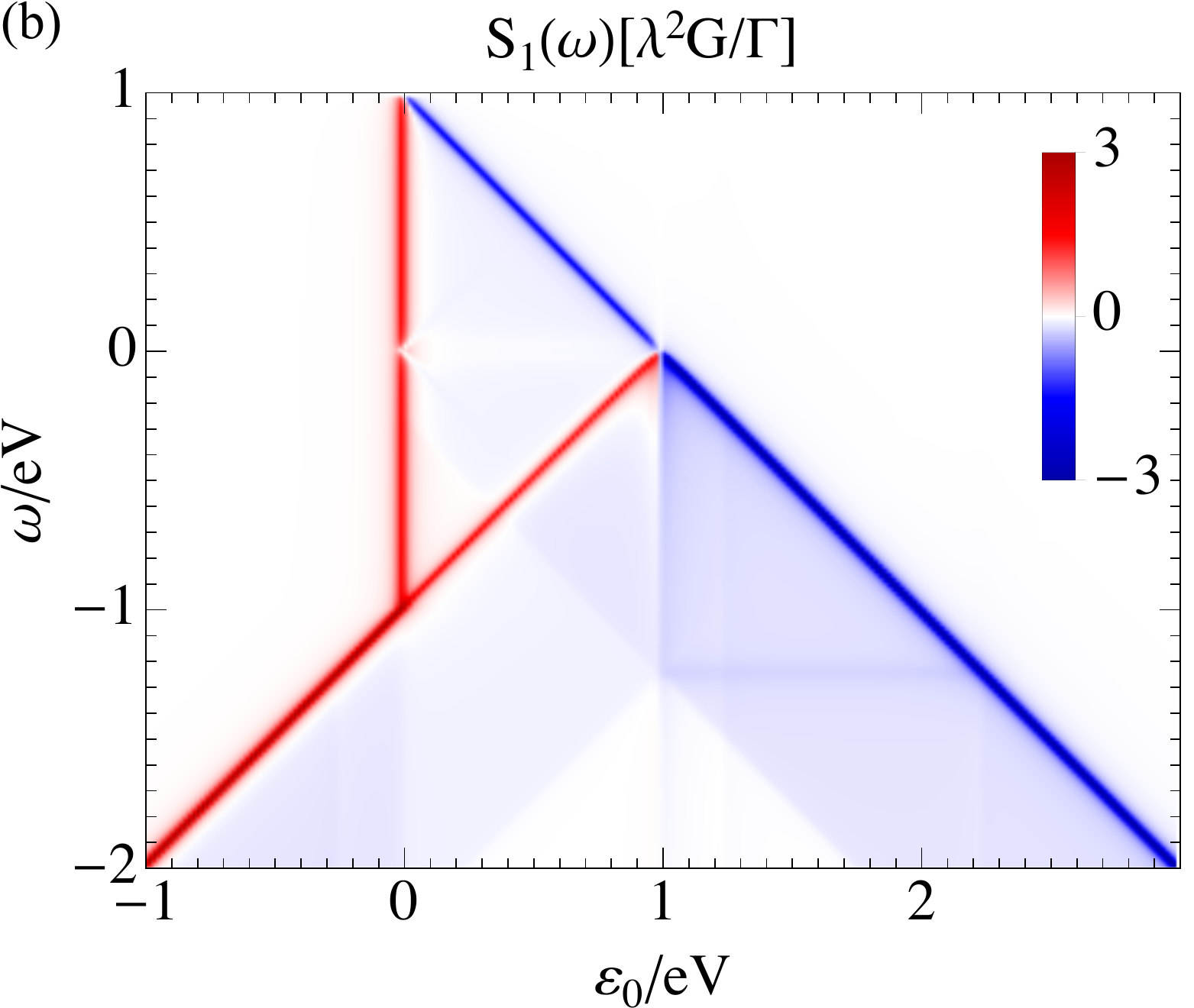}
	\end{center}
	\caption{
	Correction to the noise $S_1(\omega)$ as a function of the frequency $\omega$ and the quantum dot's energy level $\varepsilon_0$. 
	The coupling to the leads is symmetric $\Gamma_l=\Gamma_r=\Gamma=0.01eV$. 
	The  oscillator's frequency is set to (a) $\omega_0=0.25eV$ and (b) $\omega_0=1.25eV$. 	
	}  
	\label{fig:S1T} 
\end{figure}

Figure \ref{fig:S1T} shows the correction $S_1(\omega)$ to the noise as a function of the energy level $\varepsilon_0$ of the quantum dot 
and the frequency $\omega$ for the oscillator's frequency 
$\omega_0=0.25eV$ and $\omega_0=1.25eV$. 
The pattern of $S_1(\omega)$ reflects the behavior of the zero-order current noise typical 
of the resonant transport regime, as shown in Fig.~\ref{fig:S0E}.
However, the noise correction in Fig.~\ref{fig:S1T} vanishes in extended regions of the diagram as a function of  $\omega$ and $\varepsilon_0$.
As we show in Appendix \ref{sec:resonant_appendix} in Fig.~\ref{fig:S1_appendix}, 
the vanishing of $S_1(\omega)$ is related to the exact cancellation of the inelastic noise and the mean-field elastic term of the mean-field corrections. 
 
To discuss the characteristic features of the oscillator in the noise $S_1(\omega)$, we start with a oscillator frequency $\omega_0=0.25eV$ as a representative case for $\omega_0<eV$ [Fig.~\ref{fig:S1T}(a)].

Again, we consider first the case $\varepsilon_0=0$. 
In the range $|\omega|<eV$, the correction to the noise is positive and drops at $\omega=-eV$ 
to a negative value in the range $-eV<\omega<-eV-\omega_0$. 
When $\omega<-eV-\omega_0$ the correction slightly increases but remains negative. 
A qualitative argument to explain such sharp transition from a positive to a negative correction at $\omega=-eV$ is as follows.
Since the oscillator is at zero temperature, it can only absorb a vibrational energy quantum $\omega_0$.
In the range $-eV<\omega<-eV-\omega_0$, an electron is excited form the quantum dot's level at $\varepsilon_0=0$ to an energy above the left chemical potential and, 
in absence of charge-vibration interaction, it can tunnel to the left or right lead. 
However, due to the interaction, the excited electron can emit a phonon at frequency $\omega_0$ losing some energy.
After the emission, the electron has hence an energy below the left chemical potential and hence can only tunnel to the right lead. 
On the basis of the discussion for the zero-order noise in Sec.~\ref{subsec:S_0}, this explains the reduction of the noise 
and the reason why the correction results to be negative in the interval $-eV<\omega<-eV-\omega_0$ at $\varepsilon_0=0$.
Similar arguments hold at finite values of $\varepsilon_0$, where 
the noise is suppressed within the frequency range $-eV+\varepsilon_0<\omega<-eV+\varepsilon_0-\omega_0$ 
for $\varepsilon_0<eV$.  
Furthermore, a region of negative correction also appears below the line $\omega=eV-\varepsilon_0$ for $\varepsilon_0>0$ 
in the frequency range  $eV-\omega_0-\varepsilon_0<\omega<eV-\varepsilon_0$ for $0<\varepsilon_0<eV-\omega_0$ (emission noise) 
and $\varepsilon_0>eV$ (absorption noise).

Another interesting feature appears when  $\varepsilon_0$ is tuned close the left $\varepsilon_0\sim eV$  
or to the right $\varepsilon_0\sim0$  chemical potential. 
In the region delimited by $\omega<-eV+\varepsilon_0$ and $\omega<eV-\varepsilon_0$, at given frequency $\omega$, 
vibrational side bands appear in form of peaks (negative or positive)  at $\varepsilon_0= \pm \omega_0$  and 
$\varepsilon_0=eV \pm \omega_0$.
These two lateral peaks appear when the inelastic emission of a phonon is in resonance with the lateral chemical potentials. 
In this case, for instance, an electron inelastically emits a phonon by tunneling either from the level 
to the right chemical potential ($\varepsilon_0=+\omega_0$) or from the right chemical potential to the quantum dot ($\varepsilon_0=-\omega_0$). 
In both cases the energy to excite the oscillator solely comes from the applied voltage explaining the independence of the two side peaks as a function of frequency. 
A similar arguments holds when the gate voltage is tuned such that the level position is close to the left chemical potential, i.e  $\varepsilon_0 = eV$.

It is also interesting to note that in the range $eV<\varepsilon_0<eV+\omega_0$, a resonant (negative) peak appears {\sl exactly} at $\omega=-\omega_0$. 
In this case we argue that the photon energy absorbed by the whole system, quantum dot and oscillator, 
is resonant with the local vibration.

As an example of the case $\omega_0>eV$,  Figure~\ref{fig:S1T}(b) shows the correction to the noise for $\omega_0=1.25eV$.
Here, $S_1(\omega)$ has  similar features as the Fig.~\ref{fig:S1T}(a) with 
the negative correction band below the diagonal lines having now a larger width of $\omega_0=1.25eV$. 
Additionally, the lateral side peaks  are less visible compared to the previous case.

%
%
%
%
\section{Conclusion}
\label{sec:conclusions}
We study the frequency-dependent current noise in the Holstein model for a quantum dot  between two normal-conducting leads, 
in the perturbative limit which covers several realistic experimental
cases, such as single molecule junctions or suspended carbon nanotube quantum dots.
We focused on two limiting cases: the regime of energy-independent transmissions, in which we derived analytical expressions for the current noise, 
and  the regime of the resonant transport. 
Our analysis and predictions, based on analytic formulas for the
Holstein model,  constitute a reference for future studies of the frequency-dependent noise 
in tunnel junction with local charge-vibration interaction in more
complex transport situations (many conducting channels, multi-level
dots, etc.).
Finally, for the case of energy-independent transmission, we found that 
the noise induced by  the interaction of the dot's charge with the vibration 
represents the only contribution to the noise for high transmission of
the junction $T \simeq 1$. This calls for an experimental
investigation of this noise using on-chip detectors, thus providing
direct information on the vibrational states of the junction.   

\acknowledgments 
We acknowledge J.~C. Cuevas for interesting discussions and useful comments.
This research was supported by the Zukunftskolleg of the University of Konstanz and 
by the DFG through the collaborative research center SFB 767.

\appendix

\section{Keldysh Green's functions}
\label{app:GFandSE}

In this appendix, we recall the definitions and some relations of the  Green's functions 
which are used in the main text. 
We refer to the books of Refs.~[\onlinecite{Rammer:2007},\onlinecite{Cuevas-Scheer:2010}] for a detailed introduction.
The  Green's functions are defined as (we omit the spatial dependence)
\begin{align}
{G}(t,t^{\prime}) &= {G}^{11}(t,t^{\prime})=-i \langle {\mathcal{T}} \psi(t)\psi^\dagger(t^{\prime}) \rangle \, , \\
\tilde{G}(t,t^{\prime}) &= {G}^{22}(t,t^{\prime}) =-i \langle \tilde{\mathcal{T}} \psi(t)\psi^\dagger(t^{\prime}) \rangle \, ,  \\
{G}^{<}(t,t^{\prime}) &={G}^{12}(t,t^{\prime}) =i \langle \psi^\dagger(t^\prime) , \psi(t^{})  \rangle \,  , \\
{G}^{>}(t,t^{\prime}) &= {G}^{21}(t,t^{\prime}) =-i\langle \psi(t)  \psi^\dagger(t^{\prime})  \rangle  \, , \\
{G}^{R}(t,t^{\prime})  &=-i\theta(t-t^\prime) \langle \{\psi(t) , \psi^\dagger(t^{\prime}) \} \rangle  \, , \\
{G}^{A}(t,t^{\prime}) &=i\theta(t^\prime-t) \langle \{\psi(t) , \psi^\dagger(t^{\prime}) \} \rangle \, , \\
{G}^{K}(t,t^{\prime}) &= -i \langle [\psi(t) , \psi^\dagger(t^{\prime}) ] \rangle \, .\
\end{align}
with the commutator  denoted with $[\,,\,]$  and the field operator $\psi(t)$ in the Heisenberg picture.
In the case of bosonic field operators, the commutator is replaced $[\,,\,]$ by the anticommutator  $\{\,,\,\}$
and a minus sign must be added in the lesser Green's function. 

The electron Green's functions satisfy the following relations 
\begin{align}
G^R(t,t^{\prime})-G^A(t,t^{\prime})&=G^>(t,t^{\prime})-G^{<}(t,t^{\prime}) \\
G^K(t,t^{\prime})&=G^{11}(t,t^{\prime}) + {G}^{22}(t,t^{\prime}) \nonumber \\ &= G^{<}(t,t^{\prime}) + {G}^{>}(t,t^{\prime})  \\
G^{R}(t,t^{\prime})&=G^{11}(t,t^{\prime})-G^{<}(t,t^{\prime}) \nonumber \\ &=G^{>}(t,t^{\prime})-{G}^{22}(t,t^{\prime}) \\
G^{A}(t,t^{\prime})&=G^{<}(t,t^{\prime})-{G}^{22}(t,t^{\prime}) \nonumber \\ &=G^{11}(t,t^{\prime})-{G}^{>}(t,t^{\prime}) \\
G^{11}(t,t^\prime)&=G^R(t,t^\prime)+G^<(t,t^\prime) \nonumber \\ &= G^A(t,t^\prime)+G^>(t,t^\prime) \\
G^{22}(t,t^\prime)&=G^<(t,t^\prime)-G^A(t,t^\prime) \nonumber \\ &= G^>(t,t^\prime)-G^R(t,t^\prime)
\end{align}
and
\begin{align}
G^<(t,t^{\prime})&= (G^{K}(t,t^{\prime})-G^{R}(t,t^{\prime})+G^{A}(t,t^{\prime}))/2 \\
G^>(t,t^{\prime})&= (G^{K}(t,t^{\prime})+G^{R}(t,t^{\prime})-G^{A}(t,t^{\prime}))/2  
\end{align}
Further, the hermitian-conjugate of the electron Green's functions satisfy the relations
\begin{align}
{G^R(t,t^\prime)}^* &= G^A(t^\prime,t) \\
{G^<(t,t^\prime)}^* &= - G^<(t^\prime,t) \\
{G^>(t,t^\prime)}^* &= - G^>(t^\prime,t) \\
{G^{11}(t,t^\prime)}^*& = -G^{22}(t^\prime,t) \\
{G^{22}(t,t^\prime)}^* &= -G^{11}(t^\prime,t)
\end{align}
The same relations are satisfied if the electron field operators are replaced with bosonic field operators.

\section{The electron self-energy with respect to the charge-vibration interaction}
\label{app:selfenergies_vib}

We define the self-energies $\Sigma^{11}_{}$ as the following matrices
\begin{equation}
\hat{\Sigma}(\eps)=
\begin{pmatrix} \Sigma^{11}(\eps) & -\Sigma^{12}(\eps) \\ -\Sigma^{21}(\eps) & \Sigma^{22}(\eps)\end{pmatrix}
\end{equation}
with a minus sign in front of $\Sigma^{12}(\eps)$ and $\Sigma^{21}(\eps)$ due to the different position of the time arguments on the Keldysh contour. 
The leading order of the self-energy with respect to the charge-vibration coupling are proportional to $\lambda^2$ and are given by 
\begin{align}
\Sigma^{11}(\varepsilon) &\!\!=\!\! \lambda^2 \sum_s \left[\frac{1}{2}[1+n_B(\omega_0)] G^{11}(\varepsilon+s\omega_0) \right. \\ &-\left. \frac{i}{2\pi} \mathcal{P} \int d\varepsilon'  \frac{s}{\varepsilon'+s\omega_0} G^{11}(\varepsilon-\varepsilon') \right]
\\
\Sigma^{12}(\varepsilon) &\!\!=\!\!\lambda^2 n_B(\omega_0) [G^{12}(\varepsilon-\omega_0)\!\!+\!\!(1+n_B(\omega_0))G^{12}(\varepsilon+\omega_0)]
\\
\Sigma^{21}(\varepsilon) &\!\!=\!\!\lambda^2 n_B(\omega_0) [G^{21}(\varepsilon+\omega_0)\!\!+\!\!(1+n_B(\omega_0))G^{21}(\varepsilon-\omega_0)]
\\
\Sigma^{22}(\varepsilon) &\!\!=\!\! \lambda^2 \sum_s \left[ \frac{1}{2}[1+n_B(\omega_0)]G^{22}(\varepsilon+s\omega_0)  \right. \\ &+\left. \frac{i}{2\pi} \mathcal{P} \int d\varepsilon'  \frac{s}{\varepsilon'+s\omega_0} G^{22}(\varepsilon-\varepsilon') \right]
\\
\Sigma^R(\varepsilon)&\!\!=\!\! \Sigma^{11}(\varepsilon)-\Sigma^{12}(\varepsilon)
\\
\Sigma^A(\varepsilon)&\!\!=\!\! \Sigma^{12}(\varepsilon)-\Sigma^{22}(\varepsilon)
\end{align}
with the frequency $\omega_0$ of the oscillator, the Bose-distribution function $n_B(\omega)$, and the Cauchy principal value $\mathcal{P}$ of the integral. 
In the previous formulas, the electronic Green's functions $G$ are the dot's function in absence of interaction with the vibration.

\section{Real-time vertex current noise}
\label{app:vertex_realtime}
In this appendix we transform the vertex correction from the Keldysh contour time integration to the real time integration.

To this end, we first cast the vertex correction in terms of rainbow-like and tadpole-like diagrams. 
The vertex correction in Eq.~\eqref{eq:vcrealtime} can be written as
\begin{align}
S_{\mathrm{vc}}(\tau,\tau^\prime) = &\int  d\tau_1 d\tau_2 d\tau_3 \sum_{\nu=\Romannum{1},\Romannum{2}} \left\{ \right. \, \nonumber \\
&G(\tau,\tau_1)\Sigma_{\mathrm{rb}}^{\nu}(\tau_1,\tau_2)G(\tau_2,\tau_3)\Sigma_l(\tau_3,\tau)
\nonumber \\
+&\left. G(\tau,\tau_1)\Sigma_{\mathrm{tp}}^{\nu}(\tau_2,\tau_2)G(\tau_1,\tau_3)\Sigma_l(\tau_3,\tau)  \right\} \, .
\label{app:svc}
\end{align}
Here, we defined the rainbow-like and tadpole-like self-energies $\Sigma_{\mathrm{rb}}^{\nu}(\tau_1,\tau_2)$ and $\Sigma_{\mathrm{tb}}^{\nu}(\tau_1,\tau_2)$ with 
$\nu=(\Romannum{1},\Romannum{2})$ respectively. 
We remark that the energies $\Sigma_{\mathrm{rb}}^{\nu}(\tau_1,\tau_2)$ and $\Sigma_{\mathrm{tb}}^{\nu}(\tau_1,\tau_2)$ implicitly depend on the external time $\tau^\prime$ [see Eqn.~\eqref{eq:Crb1}-\eqref{eq:Ctp2}].

The difference between the self-energies labeled with $\Romannum{1}$ and $\Romannum{2}$ 
comes from the dependence of the Green's function and self-energies on the external time $\tau^\prime$. 
The rainbow-like and tadpole-like diagrams are
\begin{align}
\Sigma_{\mathrm{rb}}^{\nu}(\tau_1,\tau_2) &=  i\lambda^2 C_{\mathrm{rb}}^{\nu}(\tau_1,\tau_2)D(\tau_1,\tau_2) 
\label{eq:rb1}
\\
\Sigma_{\mathrm{tp}}^{\nu}(\tau_1,\tau_2) &= -i \lambda^2 C_{\mathrm{tp}}^{\nu}(\tau_2)D(\tau_1,\tau_2) \label{eq:tp1}
\end{align}
with the functions $C_{\mathrm{rb}}^{\nu}$ and $C_{\mathrm{rb}}^{\nu}$ given by
\begin{align}
C_{\mathrm{rb}}^{\Romannum{1}}(\tau_1,\tau_2) &= \int d\tau_4 G(\tau_1,\tau_4) \Sigma_l(\tau_4,\tau^\prime) G(\tau^\prime,\tau_2) \, ,
\label{eq:Crb1}
\\
C_{\mathrm{rb}}^{\Romannum{2}}(\tau_1,\tau_2) &= \int d\tau_4 G(\tau_1,\tau^\prime) \Sigma_l(\tau^\prime,\tau_4)G(t_4,\tau_2) \, ,
\label{eq:Ctb2}
\end{align}
and
\begin{align}
C_{\mathrm{tp}}^{\Romannum{1}}(\tau_2) &=  \int d\tau_4 G(\tau_2,\tau_4)\Sigma_l(\tau_4,\tau^\prime)G(\tau^\prime,\tau_2) \, ,
\label{eq:Ctp1}
\\
C_{\mathrm{tp}}^{\Romannum{2}}(\tau_2) &= \int d\tau_4 G(\tau_2,\tau^\prime)\Sigma_l(\tau^\prime,\tau_4)G(\tau_4,\tau_2) \, .
\label{eq:Ctp2}
\end{align}
The functions $C_{\mathrm{tp}}^{\alpha}(\tau_2)$ depend only on the time $\tau_2$ but not on $\tau_1$.

Second, we transform the vertex correction from the time on the Keldysh contour to the real time and perform a Fourier transformation. 
When transforming Eq.~\eqref{app:svc} from the contour to the real time, 
we introduce the matrix Green's functions defined in Eq.~\eqref{eq:G_el}. 
To give an example, the following  term 
\begin{equation}
A(\tau,\tau^\prime) = \int d\tau_1 G(\tau,\tau_1)\Sigma_l(\tau_1,\tau^\prime)
\end{equation}
transform as
\begin{equation}
\hat{A}(t,t^\prime) = \int dt_1 \hat{G}(t,t_1)\hat{\tau}_K \hat{\Sigma}_l(t_1,t^\prime)
\end{equation}
where we have the Pauli matrix $\hat{\tau}_K$ with $1$ and $-1$ for diagonal elements
which takes into account the position of the contour time $\tau_1$ on the Keldysh contour, see 
Ref.[\onlinecite{Rammer:2007}].
Finally, the crucial step to calculate the current noise vertex corrections is to transform the rainbow- and tadpole-like 
diagrams in Eq.~\eqref{eq:rb1} and \eqref{eq:tp1} in the real time representation  and perform a Fourier transformation. 
\begin{widetext}
The rainbow-like diagrams can be written as
\begin{align}
\hat{\Sigma}_{\mathrm{rb}}^{\nu}(\varepsilon) &= i \lambda^2 \int \frac{d\varepsilon^\prime}{2\pi} 
\begin{pmatrix}
C_{\mathrm{rb}}^{\nu,11}(\varepsilon-\varepsilon^\prime) D^{11}(\varepsilon)  & -C_{\mathrm{rb}}^{\nu,12}(\varepsilon-\varepsilon^\prime) D^{12}(\varepsilon) \\ 
-C_{\mathrm{rb}}^{\nu,21}(\varepsilon-\varepsilon^\prime) D^{21}(\varepsilon) & 
C_{\mathrm{rb}}^{\nu,22}(\varepsilon-\varepsilon^\prime) D^{22}(\varepsilon)   
\end{pmatrix} \, ,
\end{align}
with the functions 
\begin{align}
\label{eq:noisef_vortex}
\hat{C}_{\mathrm{rb}}^{\Romannum{1}}(\varepsilon)&=
\begin{pmatrix}
G^{11}(\varepsilon-\omega) & G^{12}(\varepsilon-\omega) \\ G^{21}(\varepsilon-\omega) & G^{22}(\varepsilon-\omega)
\end{pmatrix} \check{\tau}_K
\begin{pmatrix}
0& \Sigma_{l}^{12}(\varepsilon-\omega)  \\0 &  \Sigma_{l}^{22}(\varepsilon-\omega) 
\end{pmatrix}
\begin{pmatrix}
0 & 0 \\ 
G^{21}(\varepsilon)  & G^{22}(\varepsilon) 
\end{pmatrix}
\\
\hat{C}_{\mathrm{rb}}^{\Romannum{2}}(\varepsilon) &=
\begin{pmatrix}
0 & G^{12}(\varepsilon-\omega) \\ 
0 & G^{22}(\varepsilon-\omega)
\end{pmatrix}
\begin{pmatrix}
0 & 0 \\ 
\Sigma_{l}^{21}(\varepsilon) & \Sigma_{l}^{22}(\varepsilon)
\end{pmatrix} \check{\tau}_K
\begin{pmatrix}
G^{11}(\varepsilon) & G^{12}(\varepsilon) \\ 
G^{21}(\varepsilon) & G^{22}(\varepsilon)
\end{pmatrix} \, .
\end{align}
Similary, the tadpole-like diagrams are given by
\begin{align}
\hat{\Sigma}_{\mathrm{tp}}^{\nu} &= - i \lambda^2
\begin{pmatrix}  
D^R(-\omega) & 0 \\
0 &-D^A(-\omega)
\end{pmatrix} \int \frac{d\varepsilon^\prime}{2\pi}\hat{C}^{\nu,12}_{\mathrm{tp}}(\varepsilon^\prime)
\end{align}
with
\begin{align}
\label{eq:noisef_vortex}
\hat{C}_{\mathrm{tp}}^{\Romannum{1}}(\varepsilon)&=
\begin{pmatrix}
G^{11}(\varepsilon-\omega) & G^{12}(\varepsilon-\omega) \\ G^{21}(\varepsilon-\omega) & G^{22}(\varepsilon-\omega)
\end{pmatrix} \check{\tau}_K
\begin{pmatrix}
0& \Sigma_{l}^{12}(\varepsilon-\omega)  \\0 &  \Sigma_{l}^{22}(\varepsilon-\omega) 
\end{pmatrix}
\begin{pmatrix}
0 & 0 \\ 
G^{21}(\varepsilon)  & G^{22}(\varepsilon) 
\end{pmatrix}
\\
\hat{C}_{\mathrm{tp}}^{\Romannum{2}}(\varepsilon) &=
\begin{pmatrix}
0 & G^{12}(\varepsilon-\omega) \\ 
0 & G^{22}(\varepsilon-\omega)
\end{pmatrix}
\begin{pmatrix}
0 & 0 \\ 
\Sigma_{l}^{21}(\varepsilon) & \Sigma_{l}^{22}(\varepsilon)
\end{pmatrix} \check{\tau}_K
\begin{pmatrix}
G^{11}(\varepsilon) & G^{12}(\varepsilon) \\ 
G^{21}(\varepsilon) & G^{22}(\varepsilon)
\end{pmatrix} \, .
\end{align}

\section{Individual corrections $S_{\mathrm{mf}}  = S_{\mathrm{ec}}  +  S_{\mathrm{in}}$ and  $S_{\mathrm{vc}} $ for the energy-independent transmission regime.}
\label{sec:T_independent}
As discussed in the manuscript, the corrections to the noise can be divided into a mean-field elastic correction and vertex correction. 
In this appendix, we report analytic formulas for the individual contributions to the  $S_1(\omega)$ 
in the case of energy-independent transmissions. 
When $eV>\omega_0$, these corrections   are given by
	\begin{align}
	S^{eV>\omega_0}_{\mathrm{in}}(\omega) &= \frac{\lambda^2e^2 T^2}{4h\Gamma^2}  
	\begin{cases}   
   -4 (\omega + \omega_0 )& \omega  <-eV-\omega_0
	\\
	eV - 3 \omega- 3 \omega_0& -eV-\omega_0<\omega <-eV
	\\
   eV (1 - T) - (3 + T) \omega- 3 \omega_0 & -eV<\omega <\mathrm{min}(-\omega_0,-eV+\omega_0)
	\\
    eV (1 - T) - \omega- (1 - T) \omega_0 \\ \hspace{0.5cm}-\theta(eV-2\omega_0) (eV T + 2 ((1+T) \omega + \omega_0)) & \mathrm{min}(-\omega_0,\omega_0-eV)<\omega <\mathrm{max}(-\omega_0,\omega_0-eV)
	\\
   eV (1 - 2 T) -(1+ T) \omega - (1 - 2 T) \omega_0 &\mathrm{max}(-\omega_0,\omega_0-eV)<\omega<0
	\\
   eV (1 - 2 T) - (1 - T) \omega - (1 - 2 T) \omega_0 & 0<\omega<\mathrm{min}(\omega_0,eV-\omega_0)
	\\
    (-eV  + \omega_0 )T \\ +\theta(eV-2\omega_0)(eV(1-T) -(1 - 2 T) \omega - \omega_0)&\mathrm{min}(\omega_0,eV-\omega_0) <\omega<\mathrm{max}(\omega_0,eV-\omega_0) 
	\\
	(-eV + \omega )T  & \mathrm{max}(\omega_0,eV-\omega_0) < \omega < eV
	\\
	0 & \omega > eV
	\end{cases}
	\label{eq:Sin1T_Appendix}	\\
	%
	S^{eV>\omega_0}_{\mathrm{ec}}(\omega) \!&=\! \frac{\lambda^2e^2 T^2}{4h\Gamma^2}  
	\begin{cases}   
	-4(1 \!-\! 2 T) \omega \!+\! 4 T \omega_0 & \omega  \!<\!-\!eV\!-\!\omega_0
	\\
   -eV T + (-4 + 7 T) \omega + 3 T \omega_0& -eV-\omega_0<\omega <-eV
	\\
	eV (2 - 8 T + 6 T^2) - 2 \omega + 3 T (2 T \omega + \omega_0) & -eV<\omega <\mathrm{min}(-\omega_0,-eV+\omega_0)
	\\
	eV (2 \!-\! 8 T \!+\! 6 T^2) \!-\! (2 \!+\! T (1 \!-\! 4 T)) \omega \!+\! 
	2 (1 \!-\! T) T \omega_0 \\+ 
	T(2 eV T + \omega + 4 T\omega + \omega_0) \theta(
	eV - 2 \omega_0) & \mathrm{min}(-\omega_0,\omega_0-eV)<\omega <\mathrm{max}(-\omega_0,\omega_0-eV)
	\\
   2 eV (1 \!-\! 2 T)^2 \!-\! (1 \!+\! 2 T) (2 \!-\! 3 T) \omega \!+\! 
   2 (1 \!-\! 2 T) T \omega_0 &\mathrm{max}(-\omega_0,\omega_0-eV)<\omega<0
	\\
	(-1 + 2 T) (eV (-2 + 4 T) + (2 - 3 T) \omega - 2 T \omega_0) & 0<\omega<\mathrm{min}(\omega_0,eV-\omega_0)
	\\
(-1 + 2 T) (eV (-2 + 3 T) + 2 \omega - 
T (2 \omega + \omega_0) \\+ 
T(eV - 2 \omega) \theta(eV - 2 \omega_0))&\mathrm{min}(\omega_0,eV-\omega_0) <\omega<\mathrm{max}(\omega_0,eV-\omega_0) 
	\\
	(2 + T(-7 + 6 T)) (eV - \omega) & \mathrm{max}(\omega_0,eV-\omega_0) < \omega < eV
	\\
	0 & \omega > eV
	\end{cases}
	\label{eq:Sec1T_Appendix} 
	%
	\end{align}

	\begin{align}
	S^{eV>\omega_0}_{\mathrm{vc}}(\omega) &= \frac{\lambda^2e^2 T^2}{4h\Gamma^2}  
	\begin{cases}   
		2 (\omega + \omega_0) & \omega  <-eV-\omega_0
		\\
		2 (\omega + \omega_0) - T (eV + \omega + \omega_0)& -eV-\omega_0<\omega <-eV
		\\
		T (-3 + 2 T) (eV + \omega) - T \omega_0 + 
		2 (\omega+ \omega_0) & -eV<\omega <\mathrm{min}(-\omega_0,-eV+\omega_0)
		\\
		T (eV (-3 + 2 T) + \omega + 3 \omega_0 - 2 T \omega_0) +\theta(eV - 2 \omega_0)\\  
		(T (eV (-1 + 2 T) - 5 \omega + 4 T \omega - 3 \omega_0) + 
		2 (\omega + \omega_0)) & \mathrm{min}(-\omega_0,\omega_0-eV)<\omega <\mathrm{max}(-\omega_0,\omega_0-eV)
		\\
		4 T (eV (-1 + T) + 1/2 T (\omega - 2 \omega_0) + \omega_0) &\mathrm{max}(-\omega_0,\omega_0-eV)<\omega<0
		\\
		2 (-1 + T) T (2 eV - \omega - 2 \omega_0) & 0<\omega<\mathrm{min}(\omega_0,eV-\omega_0)
		\\
		2 (-1 + T) T (eV - \omega_0 + (eV - 2 \omega) \theta(
		eV - 2 \omega_0))&\mathrm{min}(\omega_0,eV-\omega_0) <\omega<\mathrm{max}(\omega_0,eV-\omega_0) 
		\\
		2 (-1 + T) T (eV - \omega)  & \mathrm{max}(\omega_0,eV-\omega_0) < \omega < eV
		\\
		0 & \omega > eV
	\end{cases}
	\label{eq:Svc1T_Appendix}	
\end{align}	
It is interesting to note, that for perfect transmission $T=1$, the emission noise  $S^{eV>\omega_0}_{\mathrm{vc}}(\omega)$ vanishes due to the factor $1-T$.
As discussed Sec.~\ref{sec:constanttransmission} the emission noise vanishes too for $\omega>eV-\omega_0$ at perfect transmission. 
From the Eqs.~\eqref{eq:Sin1T_Appendix} and \eqref{eq:Svc1T_Appendix} this can be explained by the 
exact cancellation of the two mean-field contributions,  the inelastic and the elastic term, 
in the interval $ \mathrm{max}(\omega_0,eV-\omega_0) < \omega < eV$. 

When the voltage is smaller than the frequency of the oscillator, $eV<\omega_0$ the contributions to the correction to the noise are

	\begin{align}
	S^{eV<\omega_0}_{\mathrm{in}}(\omega)  &= \frac{\lambda^2e^2}{4h\Gamma^2}  T^2
	\begin{cases}
	-4 (\omega+ \omega_0) & \omega  <-eV-\omega_0
	\\
	eV - 3 (\omega + \omega_0) & -eV-\omega_0<\omega <-\omega_0
	\\
	eV - \omega - \omega_0  & -\omega_0<\omega <-eV
	\\
	eV - \omega - \omega_0  & -eV<\omega <eV-\omega_0
	\\
	0& eV-\omega_0<\omega
	\end{cases}
	\label{eq:SEC_T2}	\\
	%
	S^{eV<\omega_0}_{\mathrm{ec}}(\omega)  &= \frac{\lambda^2e^2}{4h\Gamma^2}  T^2
	\begin{cases}
	(-4 + 8 T) \omega + 4 T \omega_0 & \omega  <-eV-\omega_0
	\\
	-eV T + (-4 + 7 T) \omega + 3 T \omega_0& -eV-\omega_0<\omega <-\omega_0
	\\
	-4 \omega + T (-eV + 5 \omega + \omega_0)  & -\omega_0<\omega <-eV
	\\
	eV (2 + T (-7 + 4 T)) + (-2 + T (-1 + 4 T)) \omega + 
	T \omega_0  & -eV<\omega <eV-\omega_0
	\\
	2 (-1 + T) (eV (-1 + 2 T) + \omega + 2 T \omega)& eV-\omega_0<\omega
	\\
	2 (-1 + T) (-1 + 2 T) (eV - \omega) & 0<\omega<eV
	\\
	0 & \omega > eV
	\end{cases}
	\label{eq:SEC_T2}	\\
	%
	S^{eV<\omega_0}_{\mathrm{vc}}(\omega)  &= \frac{\lambda^2e^2}{4h\Gamma^2}  T^2
	\begin{cases}
	2 (\omega + \omega_0) & \omega  <-eV-\omega_0
	\\
	2 (\omega + \omega_0) - T (eV + \omega + \omega_0) & -eV-\omega_0<\omega <-\omega_0
	\\
	T (-eV + \omega + \omega_0)  & -\omega_0<\omega <-eV
	\\
	T (-eV + \omega + \omega_0) & -eV<\omega <eV-\omega_0
	\\
	0& eV-\omega_0<\omega
	\end{cases}
	\label{eq:SEC_T2}	
	\end{align}

\section{Individual corrections $S_{\mathrm{mf}}  = S_{\mathrm{ec}}  +  S_{\mathrm{in}}$ and  $S_{\mathrm{vc}} $ for the resonant transport.}
\label{sec:resonant_appendix}
In this appendix we rerpot the different contributions of the correction to the noise $S_1(\omega)$ shown in Fig.~\ref{fig:S1T} 
for the case of resonant transport regime. 
\begin{figure*}[h!]
	\begin{center}		
		\includegraphics[width=0.32\columnwidth,angle=0.]{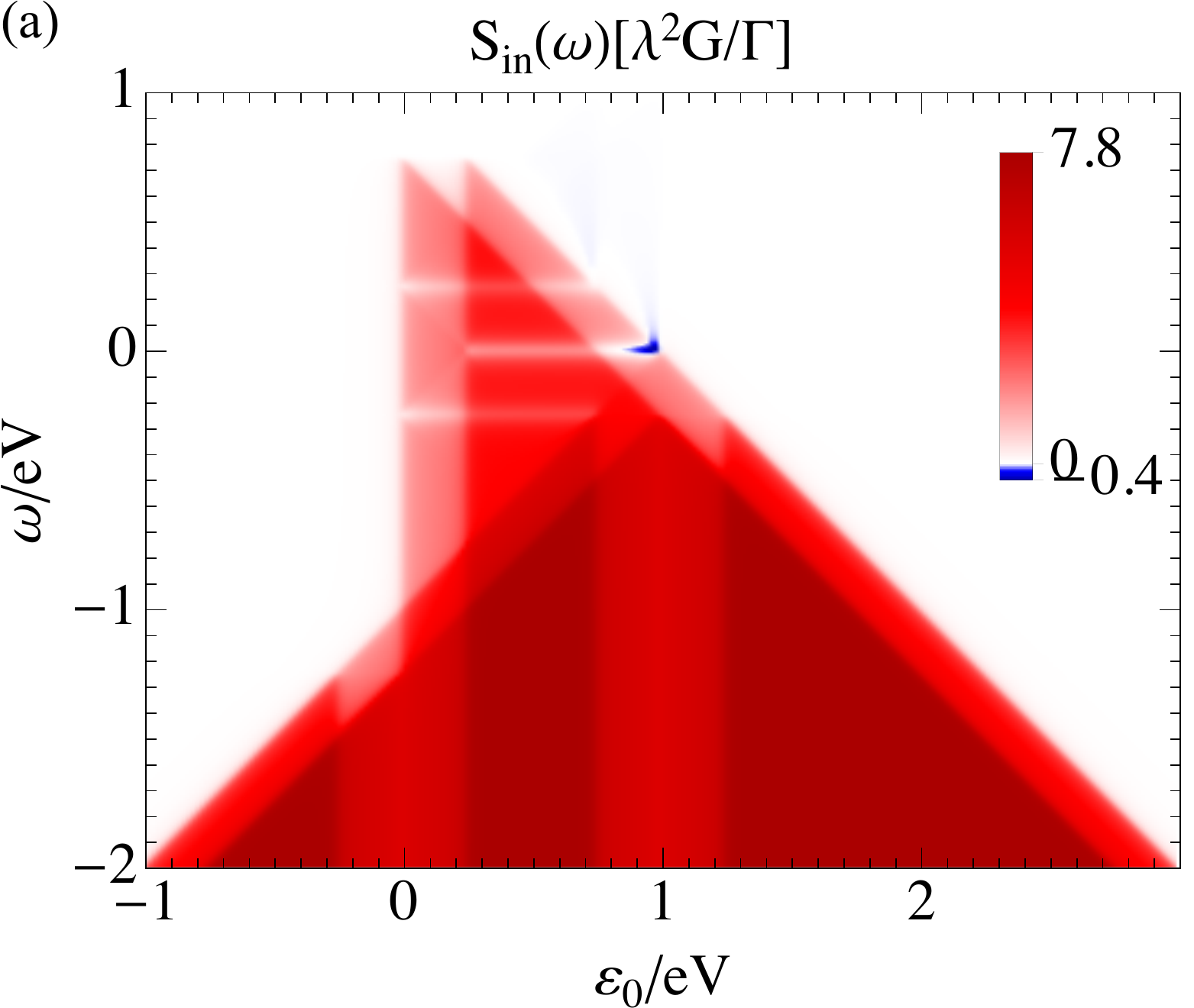}
		\includegraphics[width=0.32\columnwidth,angle=0.]{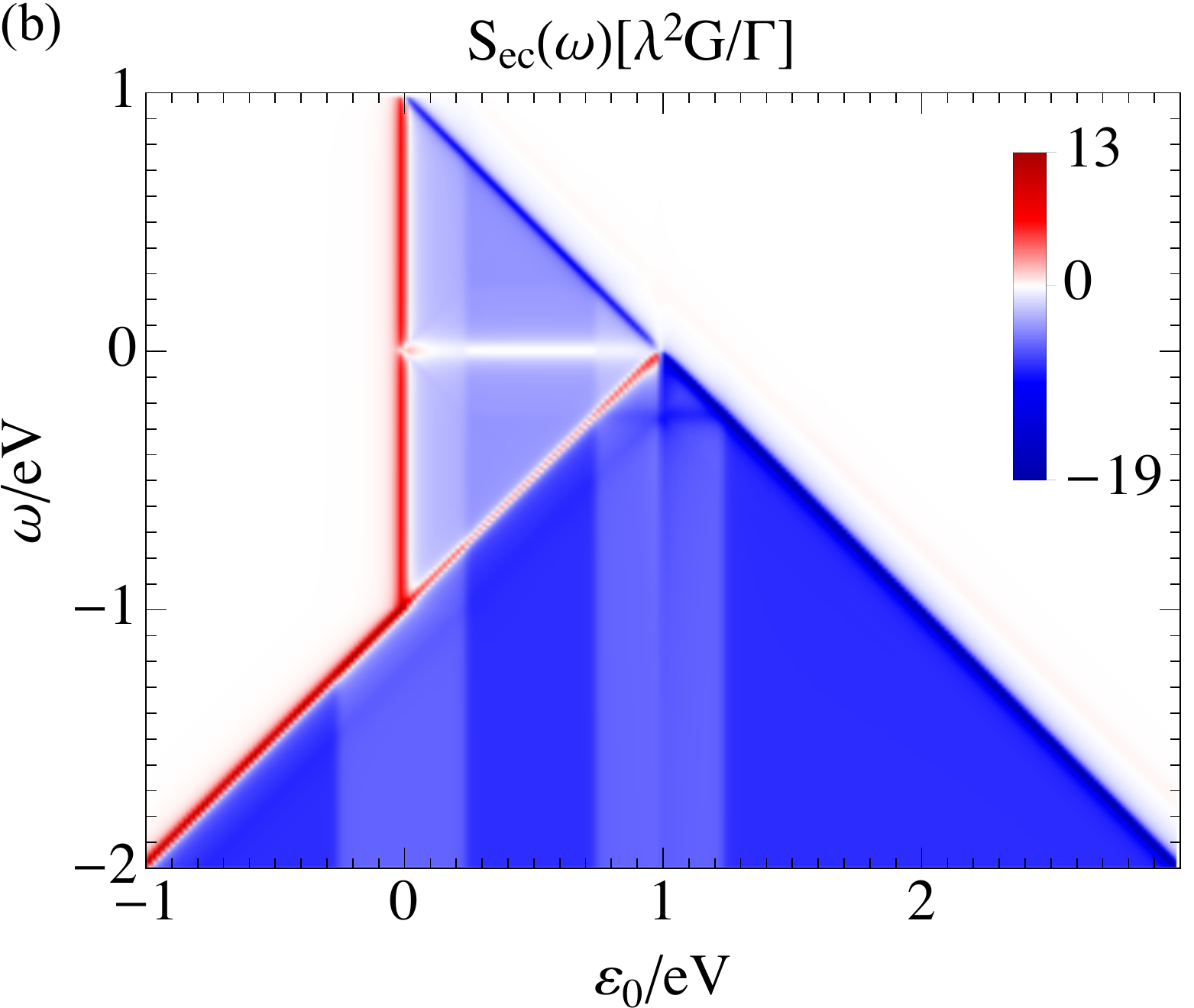}
		\includegraphics[width=0.32\columnwidth,angle=0.]{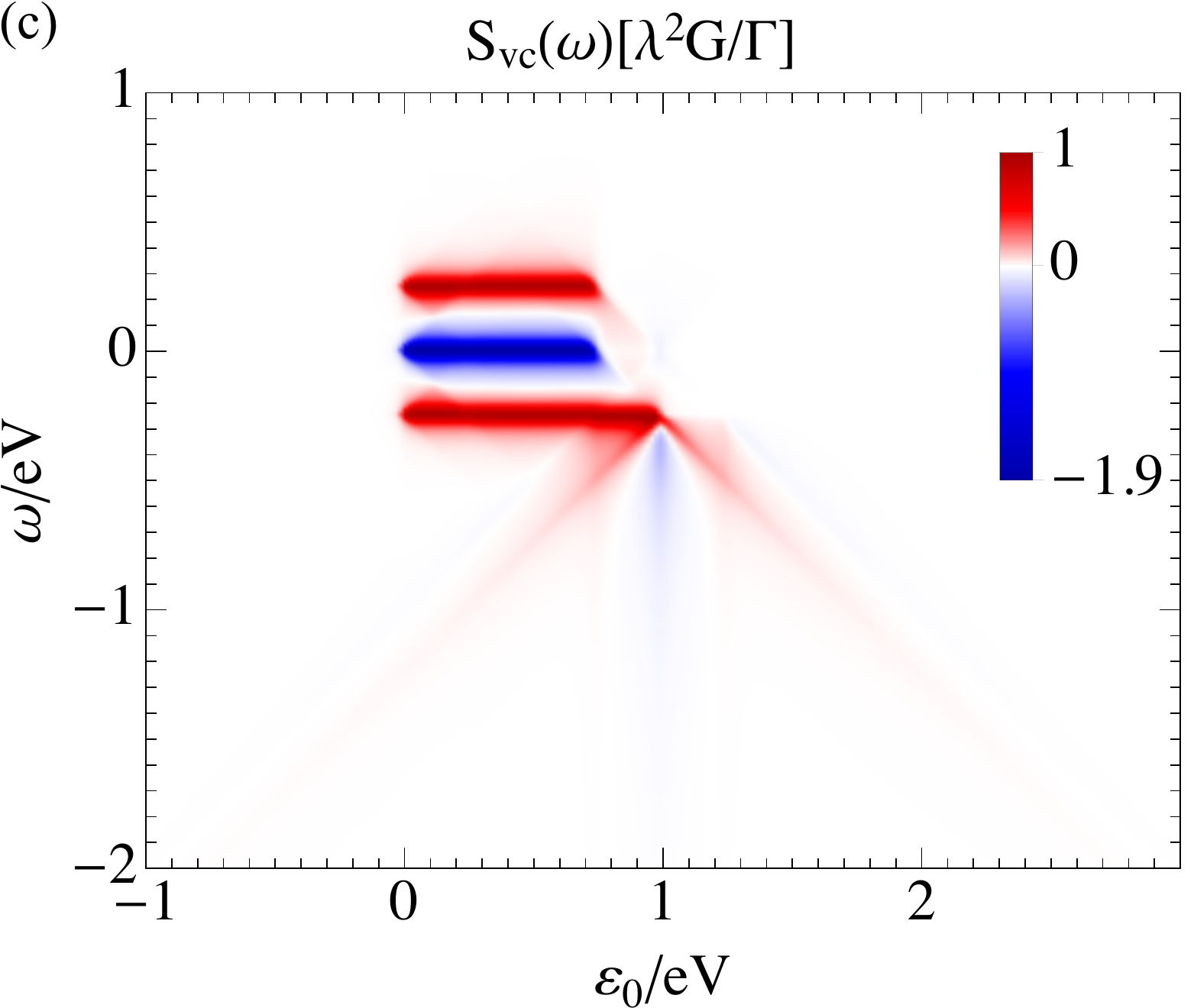}
	\end{center}
	\caption{Contributions to $S_1(\omega)$ as a function as a function of 
	the frequency $\omega$ and the quantum dot's energy level $\varepsilon_0$.
	The coupling to the leads is symmetric $\Gamma_l=\Gamma_r=\Gamma=0.01eV$ and 
	the  oscillator's frequency is set to  $\omega_0=0.25eV$.
	}  
	\label{fig:S1_appendix}
\end{figure*}

The inelastic term of the mean-field correction to the noise is reported in Fig.~\ref{fig:S1_appendix}(a). 
whereas the elastic term of the mean-field correction in Fig.~\ref{fig:S1_appendix}(b). 
Remarkably, in extended regions of the phase diagram $\omega$ and $\varepsilon_0$, 
the inelastic term and the elastic term perfectly cancel leading to a finite correction only close 
to the characteristic lines associated to the resonant transport.
These lines correspond to the stepd of the zero-order noise $S_0(\omega)$, see Fig.~\ref{fig:S0E}.
Similarly,  the vertex correction has a relevant and sharp contribution close to these lines 
and in correspondence of the vibrational  sideband at $\omega=0,\pm\omega_0$ 
in the range $0<\varepsilon_0<eV$. 
\end{widetext}

\bibliography{references}

\end{document}